\documentclass[review]{elsarticle}
\pdfoutput=1
\usepackage{lineno,hyperref}
\biboptions{sort&compress}
\usepackage{graphicx}
\graphicspath{{../pdf/}{../jpeg/}}
\usepackage{epstopdf}
\usepackage{amsmath}
\allowdisplaybreaks[4]

\newtheorem{theorem}{\bf Theorem}

\usepackage{amssymb}
\usepackage{dsfont}
\usepackage{appendix}

\makeatletter
\newif\if@restonecol
\makeatother

\usepackage[linesnumbered,ruled,vlined]{algorithm2e}
\usepackage{algpseudocode}
\usepackage{amsmath}

\usepackage{multirow}
\usepackage{xcolor}
\usepackage{colortbl}
\usepackage{array}
\usepackage{url}
\usepackage{subfig}
\usepackage{diagbox}
\usepackage{booktabs}
\usepackage{colortbl,booktabs}\definecolor{mygreen}{rgb}{.0,.70,.0}
\usepackage{amsfonts}
\usepackage{verbatim}
\usepackage{setspace}
\usepackage{changepage}
\usepackage{textcomp}
\usepackage{enumerate}
\usepackage{bm}
\usepackage{stfloats}
\usepackage{nomencl}
\usepackage{ifthen}
\renewcommand{\nomgroup}[1]{%
\ifthenelse{\equal{#1}{C}}{\item[\textit{Parameters and Constants}]}{%
\ifthenelse{\equal{#1}{V}}{\item[\textit{Variables}]}{%
\ifthenelse{\equal{#1}{S}}{\item[\textit{Sets and Indices}]}{%
\ifthenelse{\equal{#1}{A}}{\item[\textit{Abbreviations}]}{}}}}
}
\makenomenclature

\modulolinenumbers[5]

\journal{Journal of \LaTeX\ Templates}

\usepackage{multicol}

\UseRawInputEncoding

\begin{document}

\begin{frontmatter}

\title{Multi-level Coordinated Energy Management for Energy Hub in Hybrid Markets with Distributionally Robust Scheduling\vspace{-0.12cm}}
\tnotetext[mytitlenote]{This is an accepted manuscript published by Elsevier in ``Applied Energy" in 2022, available online https://doi.org/10.1016/j.apenergy.2022.118639. When citing this work, please acknowledge the original published source citation of the original paper.}

%
%
%
\author[mymainaddress,mysecondaryaddress,mytertiaryaddress]{Jiaxin Cao}
\author[mymainaddress,mysecondaryaddress,mytertiaryaddress]{Bo Yang\corref{mycorrespondingauthor}}
\cortext[mycorrespondingauthor]{Corresponding author}
\ead{bo.yang@sjtu.edu.cn}
\author[mymainaddress,mysecondaryaddress,mytertiaryaddress]{Shanying Zhu}
\author[myquaternaryaddress]{Chi Yung Chung}
\author[mymainaddress,mysecondaryaddress,mytertiaryaddress]{Xinping Guan}
\address[mymainaddress]{Department of Automation, Shanghai Jiao Tong University, Shanghai 200240, China}
\address[mysecondaryaddress]{Key Laboratory of System Control and Information Processing, Ministry of Education of China}
\address[mytertiaryaddress]{Shanghai Engineering Research Center of Industrial Intelligent Control and Management， Shanghai 200240, China}
\address[myquaternaryaddress]{Department of Electrical and Computer Engineering, University of Saskatchewan, Saskatoon SK S7N 5A9, Canada\vspace{-0.12cm}}
\begin{abstract}
Maintaining energy balance and economical operation is significant for multi-energy systems such as the energy hub (EH).
However, it is usually challenged by the frequently changing and unpredictable uncertain parameters at different timescales.
Under this scope, this paper investigates the coordinated energy management problem for day-ahead and intra-day conditions considering uncertainties of source-load and market prices concurrently.
Note that the precise knowledge of distributions about uncertainties may be unaccessible before the decision-making in day-ahead phase.
A two-stage chance-constrained model based on distributionally robust approach with ambiguous moment information is proposed to immunize scheduling strategies against the worst-case probability distributions.
The first stage is dedicated to obtaining more energy arbitrage and operation flexibility by optimizing bidding strategies in day-ahead power, natural gas and carbon trading markets.
The second stage focuses on the optimization of the worst-case expected operation cost.
It provides a robust energy equipment and load scheduling strategy for the reference of subsequent intra-day arrangements.
With respect to different variations of electrical and thermal components, an intra-day two-timescale coordination is implemented step by step.
The energy scheduling is re-dispatched circularly to minimize the operation and penalty costs.
The possible energy imbalance is also compensated by this way.
As the energy management program is nonlinear, chance-constrained and multi-stage, some linearization and dual transformation techniques are designed to enhance tractability of the program.
Experimental results show that the developed multi-level framework results in a carbon emission decrease of 37\%, and reduces energy cost averagely 3\% compared with corresponding contrasting cases.
The obtained strategy validates a good tradeoff between robustness and optimality.
\end{abstract}

\end{frontmatter}

\begin{center}{
\fbox{
  \parbox{1\linewidth}{
\begin{multicols}{2}
\begin{spacing}{1.0}
\small{
\nomenclature[Vp]{$E_{out,k},H_{out,k}$}{Available power and heat of EH through energy conversion}
\nomenclature[Vp]{$u_{e,k}(u_{g,k})$}{Purchased electricity (heat) from day-ahead markets}
\nomenclature[Vp]{$u_{c,k}$}{Carbon credit purchase margin}
\nomenclature[Vp]{$m_{e,k},m_{g,k},m_{c,k}$}{Energy budgets for electricity, heat, carbon emission}
\nomenclature[Vp]{$u_{f,k}$}{Renewable power output}
\nomenclature[Vp]{$\alpha_{d,k}$}{Dispatch factor of EH}
\nomenclature[Vp]{$D_{i,k},D_{i,k}^{u},D_{i,k}^{s}$}{Load, basic inelastic load and adjustable elastic load for $i=$ electricity (e), heat (h)}
\nomenclature[Vp]{$E_{T,k}$}{Electricity transacted in real-time power markets}
\nomenclature[Vp]{$\psi_{s,k}^{ch},\psi_{s,k}^{dis},\psi_{s,k}$}{Charged/discharged power and energy of storage $\psi=$ electrical (E), heat (H)}
\nomenclature[Vp]{$\xi_{k}$}{Uncertain variables}
\nomenclature[Vp]{$\bm{x}$}{Shorthands of decision variables}
\nomenclature[Vp]{$p_{c}^{k}$}{Real-time power market price}
\nomenclature[Vp]{$r_{k},Q_{k},w_{k},y_{k}$}{Auxiliary variables in DRO approach}
\nomenclature[Cp]{$\eta_{*}$}{Conversion efficiency of unit *}
\nomenclature[Cp]{$p_{ie,k}(p_{ig,k})$}{Day-ahead electricity (heat) prices}
\nomenclature[Cp]{$MN_{k}$}{Disposable energy budgets}
\nomenclature[Cp]{$D_{i}^{s,day},D_{i}^{r}$}{Minimum daily elastic load and drop-off/pick-up rate for $i=$ electricity (e), heat (h)}
\nomenclature[Cp]{$(\cdot)^{\max/\min}$}{Maximum/minimum value of $(\cdot)$}
\nomenclature[Cp]{$\varepsilon$}{Confidence level of chance constraints}
\nomenclature[Cp]{$a_{s,k}^{E},a_{s,k}^{H}$}{Cost parameters of storages}
\nomenclature[Cp]{$\alpha_{i,k},\beta_{i,k}$}{Utility coefficients for load $i=$ electricity (e), heat (h)}
\nomenclature[Cp]{$K_{c,k},K_{H,k}$}{Carbon trading price and over-emission penalty price}
\nomenclature[Cp]{$C_{o2e},C_{o2g}$}{Carbon intensity in power and gas links}
\nomenclature[SO]{$k,m$}{Index of the time slot}
\nomenclature[SO]{${{\mathcal{P}}_{k}}$}{Ambiguity sets}
\nomenclature[SO]{${{\mathbb{E}}_{{{\mathbb{P}}}}[\cdot]}$}{Expectation with respect to distribution ${\mathbb{P}}$}
\nomenclature[SO]{${\mathbb{P}}_{k}\{\cdot\}$}{Probability}
\nomenclature[AO]{EH}{Energy hub}
\nomenclature[AO]{MESs}{Multi-energy systems}
\nomenclature[AO]{PDF}{Probability distribution function}
\nomenclature[AO]{SP}{Stochastic programming}
\nomenclature[AO]{RO}{Robust optimization}
\nomenclature[AO]{DRO}{Distributionally robust optimization}
\nomenclature[AO]{CC}{Chance-constrained}
\nomenclature[AO]{RESs}{Renewable energy sources}
\printnomenclature[0.71in]
}
\end{spacing}
\end{multicols}
  }
}
}
\end{center}

\section{Introduction}
\subsection{Background and Motivation}
With the concern of energy crunch and environmental deterioration, multi-energy systems (MESs)~\cite{wang2018modeling} are acknowledged to be a promising approach to increasing the efficiency of energy utilization by exploiting 
complementary (alternative) energy sources and storages.
Energy hub (EH) \cite{mohammadi2017energy}, as one popular conceptual model of MESs via mapping multi-energy conversions to input-output ports, shows favorable performance for improving the penetration of renewable energy sources (RESs) and accelerating the low-carbon transition.
However, as more renewables penetrate into the energy system, despite yielding low-cost and environmental welfare, their highly uncertain characteristics challenge the system control with significant impacts on secure and economical operations \cite{davatgaran2018optimal}.
Moreover, since diverse variable natures of energy loads in EH, it is necessary to develop applicable energy management programs in presence of uncertainties for the optimal EH operation issue.

In the context of MESs as aforementioned, the EH has to deal with multi-energy carriers in order to minimize the overall system cost and cut down carbon intensity. Moreover, to provide more operation feasibility in load management and transaction, in addition to power markets, EH also participates in other energy markets such as natural gas and carbon trading markets which are generally referred to as hybrid markets.
It is noteworthy that, different from the physical electricity and natural gas markets, carbon trading system \cite{weng2018review} as a special virtual energy market aims at reducing carbon emissions and relieving the burden of climate warming indirectly by the price lever.
For instance, with the purpose of removing 55\% of carbon emissions at least by 2030, European Union emission trading system \cite{zhou2019carbon} adopts the cap-and-trade mechanism which enables energy entities to trade their holdings of emission credits (permits or allowances) by taking account of carbon emissions.
Similarly following rules of cap-and-trade, in China, the national carbon market has been launched on July 16, 2021 \cite{carbon716}, where carbon credits are bought/sold as a commodity
to motivate entities implementing low-carbon behaviors.
On the other hand, from the aspect of time-span, the market form usually involves ``day-ahead + real-time", ``day-ahead + balancing", ``day-ahead + intra-day + balancing", etc.,
taking electricity market for example \cite{conejo2010decision}.
Although the type of electricity market may vary worldwide, it has similar market structures. In this paper, to present the work as comprehensive and brief as possible, we mainly focus on the
``day-ahead + real-time" form of electricity markets, day-ahead natural gas market and carbon trading market considering their different fluctuating natures.

Against this backdrop, many factors complicate the decision-making of EH energy management.
Decisions of energy scheduling and bidding in power, gas and carbon markets are dependent.
Besides, energy bidding decisions in day-ahead markets are usually made before more accurate information of renewable generation, demand and market prices are available.
Therefore, how to minimize total operation costs of EH via strategies of bidding-scheduling in a hybrid market environment considering different operation properties with granular regulation hedging against multiple uncertainties
is a crucial problem to be addressed, which is further explored in this paper.
\subsection{Literature Review}
In the presence of renewable generation and energy storage deployments, making the best energy management schedule of the modern energy system
is an intricate optimization program for the economic and reliable operation,
and becomes more complicated by taking into account total operation cost and energy imbalance factors~\cite{azizivahed2017new}.
Recognizing the significance and opportunity of energy management for EH under different uncertainties,
an amount of literature has been carried out and made considerable contributions.
The relevant literature can be sectionalized into the following three parts: two-stage probabilistic optimization of EH scheduling, distributionally robust energy management, multi-timescale coordination strategy.
\subsubsection{Two-stage probabilistic optimization of EH scheduling}
To deal with operation issues under demand uncertainties, \cite{yong2021day} indicates that the energy management can be done more efficiently by a more precise day-ahead dispatch plan
based on the probabilistic methods considering predictable information updates.
Different from the single optimization in day-ahead phase, some studies have also been conducted employing a two-stage probabilistic optimization framework to jointly optimize the day-ahead and real-time energy dispatches with different markets.
Authors in \cite{al2021two} investigate
a phase-based decision-making model
for the participation of wind generation owners in different power markets based on a two-stage stochastic programming (SP) method and exhibit the importance of information updating in sequential optimization.
A two-stage SP with chance constraints framework is established in \cite{zhao2018strategic}
by
scenario trees to obtain a tradeoff between costs and service quality which manages both the day-ahead bidding and real-time dispatch strategy of EH incorporating AC/DC microgrids.
Based on robust optimization (RO) method, the uncertain power market price is considered in \cite{7307233} via taking values in given intervals. An optimal bidding strategy in day-ahead and real-time markets is obtained for loads, wind plant and energy storages.
Ref. \cite{aboli2019joint} proposes a two-stage RO for microgrid operation with a mixed-integer linear programming model at the first day-ahead stage and a full robust model at the second real-time stage.
Results are obtained based on Benders decomposition methods and show that energy management
plans are more profitable via joint optimization in two stages.
Both these two-stage SP and RO studies have shown significant technical benefits in the decision-making of energy management under uncertainties.
However, it is noted that, in practice,
to obtain precise probability distribution functions (PDFs) for SP, massive scenarios are required \cite{chen2018analyzing} which tends to be computationally demanding and restricted to insufficient available data.
Despite the advantage of distribution-free models, RO methods are adopted to obtain energy scheduling solutions according to the worst-case scenario \cite{parisio2012robust} and may show more conservative than SP.
\subsubsection{Two-stage distributionally robust energy management}
In this context, inheriting the strengths of both SP and RO aspects, some two-stage distributionally robust optimization (DRO) studies can elaborate PDFs in functional uncertain sets \cite{8254387}, and have attracted extensive attention recently in energy management \cite{wang2020wasserstein}.
For example, Ref. \cite{zhao2020economic} has designed a two-stage DRO model with distance-based distribution ambiguity sets to capture renewable uncertainties,
which procures both day-ahead preparations and real-time adaptive regulations for the volt-pressure optimization of EH.
In \cite{zhou2019distributionally}, a co-optimization approach for energy and reserve dispatch is proposed based on the two-stage DRO where renewable and ambient temperature uncertainties are described in exact known moment-based ambiguity sets.
Hu et al. \cite{hu2020distributionally}  develop a similar two-stage DRO model to study the short-term trading strategies for waste-to-energy with a combined heat and power (CHP) plant in day-ahead power markets.
The literature review suggests that most explorations have focused on the strict energy balance in the  day-ahead scheduling.
However, issues of dynamic fluctuations in renewable outputs and different variations in power-heat demands further complicate the real-time operation, which
are not well addressed in the real-time dispatch.
\subsubsection{Multi-timescale coordination strategy}
Considering different variation characters of electrical-thermal segments,
some studies related to the multi-timescale coordination energy management have been done.
A decentralized energy management model with the multiple timescale framework is proposed for integrated EHs \cite{li2018event} where the renewable and load uncertainties are accommodated during different periods.
In \cite{gu2017online}, making up for prediction errors of source-load in a long-term period, a rolling optimization with minutes timescale is utilized to adjust day-ahead scheduling plans and optimize the intra-day operation costs.
Notably, these multi-timescale scheduling works do not take account of uncertainties in day-ahead stages 
which tend to produce a less robust dispatch.
It is because deterministic day-ahead models are not enough to ensure the feasibility of referrals for any real-time scenarios.
To adapt uncertainties in rolling optimization, Ref. \cite{9364730} combines the scenario-based robust optimization with receding horizons which maximizes the revenue of a virtual power plant in power and reserve markets.
Day-ahead schedules are settled by a stochastic problem, and power variations are penalized in two close-to-real-time dispatches.
Similar multi-timescale schemes have been presented in problems of investment planning \cite{liu2018multistage} and the portfolio \cite{wang2021multi} for a CHP plant.
In consideration of difficulties to cover all the underlying scenarios with SP, in \cite{8454320}, a novel framework of multi-timescale rolling optimization with day-ahead DRO scheduling and intra-day adjustment is introduced to compensate prediction deviations in renewables and load for an AC/DC hybrid microgrid.
Nevertheless, models do not consider uncertainties of market prices \cite{li2018event,gu2017online,8454320} and ambiguity PDFs \cite{li2018event,gu2017online,9364730,liu2018multistage,wang2021multi} which may influence the robustness and profits of MESs. In addition, the gas market \cite{9364730,8454320} and carbon trading market \cite{li2018event,gu2017online,9364730,liu2018multistage,wang2021multi,8454320} are also not included. 
However,
few studies found have some distinguished contributions in combined energy bidding strategies of EH within carbon markets \cite{9174952,coelho2021network}. They mainly focus on a single period dispatch (day-ahead or intra-day) and do not involve uncertain parameters and probability distributions.

Table~\ref{Tab22} has provided a summarized comparison among the relevant literature mainly discussed in this section.
It is found that the relevant DRO-based coordinated energy management approach supporting the bidding-scheduling in hybrid energy trading markets considering the multi-timescale character of demand
remains insufficient in the literature.
Indeed, ambiguous PDFs of different uncertainties and distinct varying features of power-heat load will further complicate the integrated energy management and usually result in computationally intractable.
Moreover, the carbon emission and credit trading are coupled with power and gas consumption processes. EH has more difficulties in decision-making to purchase or sell energy as well as how to make sound dispatch strategies for different energy resources before giving access to precise information of uncertainties.
To address the knowledge gap identified in this section, the issues of joint optimization in day-ahead and intra-day energy management for EH considering cooperative participation in hybrid energy and carbon markets and uncertain PDFs are investigated.
The major contributions of this work are summarized as threefold:
\begin{itemize}
\item[1)]
A multi-level scheduling framework of EH is developed to obtain more flexibility and competitiveness by energy arbitrage in hybrid day-ahead power, gas and carbon trading markets and real-time power market with coordinated energy management in day-ahead and intra-day periods.
It jointly computes the energy bidding, energy conversion and demand regulation in presence of uncertainties of real-time market price, renewable output and inelastic power load.
\item[2)] A chance-constrained (CC) two-stage DRO model is designed for the day-ahead schedule of EH to hedge against the worst-case probability distributions of uncertainties.
    Comparatively to the two-stage SP, this model avoids the utilization of prior precise knowledge of PDFs and calculation challenges in scenario depiction.
    Besides, it further addresses the conservative of two-stage RO.
    Meanwhile, chance constraints are imposed to restrict the probability of power supply shortage.
    Specifically, reformulation methodologies are also proposed by utilizing the linearization technique and duality theory
    to tackle the problem in a tractable manner.
\item[3)] Accommodating distinct variations of electrical and thermal components,
a two-timescale coordination mechanism is proposed for the intra-day dispatch allowing decisions to be re-optimized circularly
based on available updated information at different time periods.
Connections of power storage and demand in the two-timescale scenario are further imposed to minimize the operation and penalty costs while ensuring energy balance and operation flexibility in the cyclic dispatch under any realization of uncertainties.
\end{itemize}
\definecolor{mygray}{gray}{.9}
\newcommand{\tabincell}[2]{
\begin{tabular}{@{}#1@{}}#2\end{tabular}
}
\begin{table}[t]
\centering
\fontsize{6.5}{6}\selectfont
\caption{Comparisons between related literature.\vspace{-0.15cm}}
\setlength{\tabcolsep}{2.2mm}{%
\begin{tabular}{ccccccc}

\toprule
\rowcolor{mygray}Ref. &\tabincell{c}{Type of market}&\tabincell{c}{DA dispatch}&\tabincell{c}{ID dispatch}&\tabincell{c}{Multi-timescale}&\tabincell{c}{Uncertain parameters}&\tabincell{c}{Problem type}\\[2pt] \midrule
\rowcolor{mygray}\cite{zhao2018strategic}&\tabincell{c}{DA and RT\\power market}&  Yes &  Yes &  No & \tabincell{c}{Non-schedulable sources,\\loads, market price} &\tabincell{c}{Two-stage SP with\\chance constraints}\\[8pt]
\cite{aboli2019joint}&\tabincell{c}{DA and RT\\power market}&  Yes &  Yes &  No & \tabincell{c}{Renewable generation,\\market price}&Two-stage RO\\[8pt]
\rowcolor{mygray}\cite{hu2020distributionally}&\tabincell{c}{DA\\power market}& Yes &  Yes &  No & \tabincell{c}{Source supply,\\market price, demand} &Two-stage DRO\\[8pt]
\cite{li2018event}&\tabincell{c}{DA power and\\gas market}& Yes &  Yes &  Yes &  Renewable generation &  \tabincell{c}{Deterministic\\optimization}\\[8pt]
\rowcolor{mygray}\cite{wang2021multi}&\tabincell{c}{DA power and\\gas (heat) market,\\RT power market}&  Yes &  Yes &  Yes &\tabincell{c}{Wind generation,\\demand, market price}&Two-stage SP\\[8pt]
\cite{8454320}&\tabincell{c}{DA\\Power market}& Yes &  Yes &  Yes &Source-load &Two-stage DRO\\[8pt]
\rowcolor{mygray}\cite{coelho2021network}&\tabincell{c}{DA power, gas, \\and carbon market}& Yes &  No &  No &  No &  \tabincell{c}{Deterministic\\optimization}\\[8pt]
\tabincell{c}{This\\work}&\tabincell{c}{DA power, gas and\\carbon market,\\RT power market}& Yes & Yes & Yes &\tabincell{c}{Renewable generation,\\demand, market price}&\tabincell{c}{Two-stage DRO with\\chance constraints}\\[8pt] \bottomrule
\end{tabular}%

}
\label{Tab22}
\tiny{DA: day-ahead, RT: real-time, ID: intra-day, SP: stochastic programming, RO: robust optimization, DRO: distributionally robust optimization}\\
\vspace{-0.38cm}
\end{table}

The rest of this paper is organized as follows.
In Section~\ref{sec2},  we present the multi-level energy management architecture. The mathematical model of EH is given in Section~\ref{sec221}.
After that we formulate the optimization problem of day-ahead and intra-day periods in Sections~\ref{sec24} and \ref{sec25} respectively,
which are further processed in Section~\ref{sec3} with a corresponding operation flowchart.
The simulation results with analyses are provided in Section~\ref{sec4}. Finally, conclusions are given in Section~\ref{sec5}.
\section{Energy Management Architecture}\label{sec2}
\begin{figure}[ht]
\centering
\includegraphics[height=11.7cm,width=15.2cm]{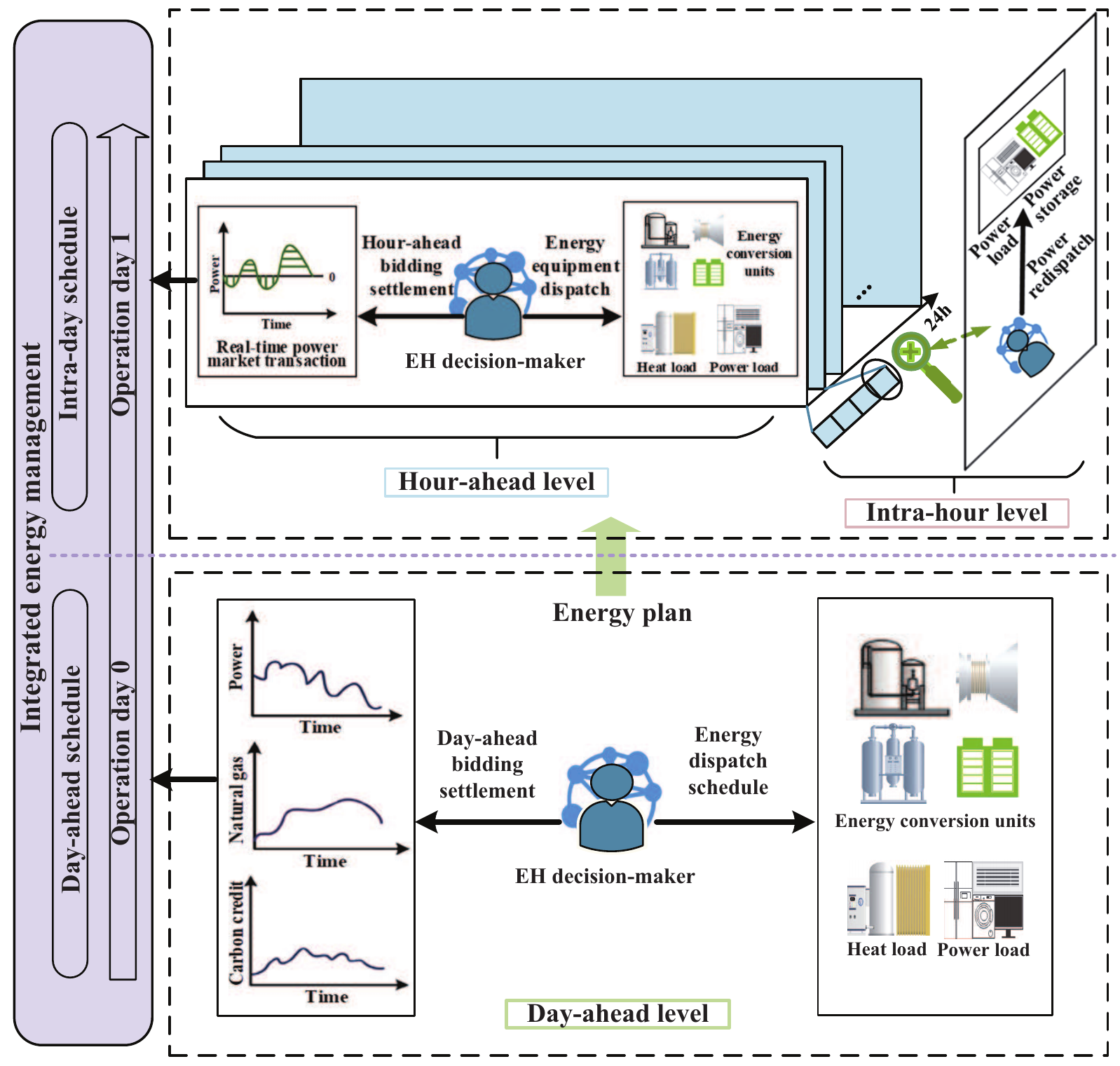}
\caption{Multi-level energy management architecture for EH.}
\label{fig3}
\vspace{-0.22cm}
\end{figure}
This section provides an overview of the proposed multi-level energy management structure as presented in Fig.\ref{fig3}.
It aims to minimize the operation cost and energy transaction cost in hybrid markets (e.g. day-ahead power, natural gas and carbon markets and real-time power market) which run simultaneously and sequentially.
In practice, day-ahead energy markets enable EH to make a commitment to purchase power, gas and carbon credit a day before the operation day which is helpful in avoiding the volatility of energy prices.
Real-time power markets allow EH to balance energy mismatch between day-ahead procurement and fluctuant load by purchasing or selling energy in the course of operation day.
And in all markets, the EH is assumed to be a price-taker \cite{8701457} considering the relatively small trading scale. In other words, the schedule operation does not affect market prices.
Considering the inherent temporal sequences of markets and electric-thermal variation characteristics, the main idea of the multi-level integrated energy management is to optimize energy utilisation
in day-ahead and intra-day horizons cooperatively including the transaction settlement in hybrid markets, dispatch in energy conversion units and load schedule.
The coordination in different level schedules is introduced briefly as follows.
The detailed discussions and procedures are described in later subsequent sections.

1) \emph{Day-ahead level:}
    The EH schedules energy transactions in each market and provides a referential management plan for energy equipment and demand ahead of the next operation day. The scheduling strategy is obtained once daily through a two-stage DRO model considering the worst-case uncertain PDFs.
    Specifically, the first-stage decisions related to energy bidding in day-ahead electricity, natural gas and carbon markets are settled without prior knowledge of uncertainties, i.e., the actual real-time market prices, renewable generation and power demand. In contrast, the second-stage decisions such as the volume of energy to sell or purchase in real-time power market, the energy conversion and demand schedules are reliant on the available information of aforementioned uncertainties.

2) \emph{Hour-ahead level:} The real-time power market price of the upcoming hour is unveiled step by step, and the obtained information regarding stochastic supply-demand is more accurate.
Based on this updated information and the day-ahead referential schedule, energy management strategies of EH components are re-optimized in a receding way.
The obtained optimal strategy of the first time slot can be also used for reference in the corresponding intra-hour level.

3) \emph{Intra-hour level:} Considering the intrinsic different timescales of electrical components and thermal components, a fine-grained model is proposed to revise the electricity consumption dependent on the optimized day-ahead and hour-ahead plans. Finally, the unbalanced power can be absorbed.
\section{Energy Hub Model Development}\label{sec221}
\begin{figure}[htbp]
\centering
\includegraphics[width=3.2in]{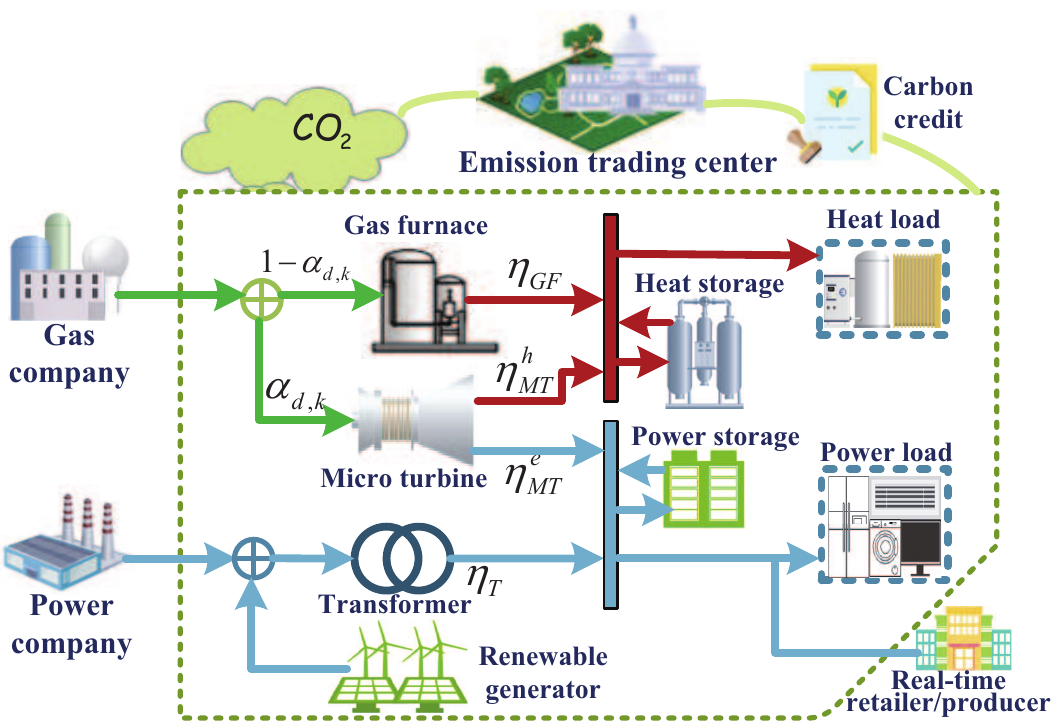}
\caption{Schematic of the proposed EH model.}
\label{fig1}
\vspace{-0.22cm}
\end{figure}

The structure of EH is shown as Fig.\ref{fig1}. The energy conversion devices in EH mainly include gas furnace with the efficiency parameter $\eta_{G}$, micro turbine with gas-to-heat and gas-to-electricity efficiencies $\eta_{M}^{h}$ and $\eta_{M}^{e}$, renewable generators, transformer with the coefficient $\eta_T$, battery and heat storages. The energy conversion constraints of EH in time slot $k$ are described as\footnote{Note that the following input-output format can be extended to formulate nonlinear power-heat production relationship by including variable energy conversion ratios. However, due to nonlinear operating conditions \cite{yong2021day}, some piecewise linearization and non-convex model should be coped with, which will complicate problem formulations and may require some meta-heuristic optimization methods. This direction will not be elaborated on here.}
\begin{eqnarray}
&&{{E}_{out,k}}={{\eta }_{T}}({{u}_{e,k}}+{{u}_{f,k}})+{{\alpha }_{d,k}}\eta _{M}^{\text{e}}{{u}_{g,k}}-{{E}_{s,k}},\
\label{eq1}\\
&&{{H}_{out,k}}=\left[ {{\alpha }_{d,k}}\eta _{M}^{h}+(1-{{\alpha }_{d,k}}){{\eta }_{G}} \right]{{u}_{g,k}}-{{H}_{s,k}},\
\label{eq2}
\end{eqnarray}
where $\alpha_{d,k}$ represents the dispatch ratio of natural gas into micro turbine and gas furnace. $E_{s,k}$ and $H_{s,k}$ denote the power and heat filling into (positive) or released from (negative) storages in per slot.
We use $u_{f,k}$ to denote the uncertain renewable generation in EH.
To meet users' energy demands, $E_{out,k}$ and $H_{out,k}$ are the available power and heat of EH through energy conversion with the purchased electricity $u_{e,k}$ and natural gas $u_{g,k}$ from utility companies in day-ahead energy markets.
Furthermore, the EH is assumed to have disposable energy budgets $MN_{k}$, which is allocated to purchase power $m_{e,k}$, natural gas $m_{g,k}$ and carbon emissions rights $m_{c,k}$ with the following constraints:
\begin{eqnarray}
&&{{m}_{e,k}}+{{m}_{g,k}}+{{m}_{c,k}}\le M{{N}_{k}},\
\label{eq3}\\
&&{{m}_{e,k}},\;{{m}_{g,k}},\;{{m}_{c,k}}\ge 0,\
\label{eq4}\\
&&{{u}_{e,k}}=\frac{{{m}_{e,k}}}{{{p}_{ie,k}}},\quad {{u}_{g,k}}=\frac{{{m}_{g,k}}}{{{p}_{ig,k}}},\
\label{eq5}
\end{eqnarray}
where $p_{ie,k}$ and $p_{ig,k}$ denote day-ahead energy prices which are settled by the power and natural gas companies beforehand.

In EH, the local users' electricity demand $D_{e,k}$ and heat demand $D_{h,k}$ are modeled by taking account of the basic inelastic load $(D_{e,k}^{u}, D_{h,k}^{u})$ and adjustable elastic load $(D_{e,k}^{s}, D_{h,k}^{s})$, which are constrained by
\begin{eqnarray}
&&{{D}_{e,k}}=D_{e,k}^{u}+D_{e,k}^{s},\
\label{eq6}\\
&&{{D}_{h,k}}=D_{h,k}^{u}+D_{h,k}^{s}.\
\label{eq7}
\end{eqnarray}
Besides, the elastic load $D_{i,k}^{s}, (i\in\{e, h\})$ over a time horizon $T\!=\!24$h also satisfies the minimum daily consumption level $D_{i}^{s,day}$. In each slot $\Delta k$, it is regulated within an interval $[D_{i,k}^{s,\min }, D_{i,k}^{s,\max }]$ and imposed by ramp constraints with the drop-off/pick-up rate $D_{i}^{r}$ ($i\in\{e, h\}$), i.e.,
\begin{eqnarray}
&&\sum\nolimits_{k=1}^{T}{D_{i,k}^{s}}\ge D_{i}^{s,day},\
\label{eq8}\\
&&D_{i,k}^{s,\min}\Delta k\le D_{i,k}^{s}\le D_{i,k}^{s,\max}\Delta k,\
\label{eq9}\\
&&-D_{i}^{r}\Delta k\leq D_{i,k+1}^{s}\!-\!D_{i,k}^{s}\leq D_{i}^{r}\Delta k.\
\label{eq10}
\end{eqnarray}

As for energy conversions in micro turbine and gas furnace, they are imposed by the dispatch operation, minimum production and maximum capacity limitations of micro turbine $G_{MT}^{\max}$, and gas furnace $G_{GF}^{\max}$, i.e., $0\le {{\alpha }_{d,k}}\le 1$, $0\le {{\alpha}_{d,k}}{{u}_{g,k}}\le G_{MT}^{\max}$ and $0\le (1-{{\alpha }_{d,k}}){{u}_{g,k}}\le G_{GF}^{\max}$.
Combining the heat supply-demand relationship $H_{out,k}=D_{h,k}$ with conversion conditions \eqref{eq2}, \eqref{eq5}, \eqref{eq7}, the above operation constraints can be converted into
\begin{equation}
\left\{ \begin{aligned}
&{{p}_{ig,k}}(D_{h,k}^{u}+D_{h,k}^{s}+{{H}_{s,k}})-{{\eta }_{G}}{{m}_{g,k}}\le 0, \\
&{{m}_{g,k}}\eta _{M}^{h}-{{p}_{ig,k}}(D_{h,k}^{u}+D_{h,k}^{s}+{{H}_{s,k}})\le 0, \\
&{{\eta }_{G}}{{m}_{g,k}}-{{p}_{ig,k}}(D_{h,k}^{u}+D_{h,k}^{s}+{{H}_{s,k}})-G_{MT}^{\max }{{p}_{ig,k}}({{\eta }_{G}}-\eta _{M}^{h})\le 0, \\
&{{p}_{ig,k}}(D_{h,k}^{u}+D_{h,k}^{s}+{{H}_{s,k}})-{{m}_{g,k}}\eta _{M}^{h}-G_{GF}^{\max }{{p}_{ig,k}}({{\eta }_{G}}-\eta _{M}^{h})\le 0. \\
\end{aligned} \right.\
\label{eq11}
\end{equation}
Note that binary variables related to startup and shutdown operations with nonlinear constraints could make the energy market non-convex \cite{9174952}.
On the other hand, as mentioned in \cite{hu2020distributionally}, usually these energy conversion devices run continuously, and few
units shut down for idle.
Since our work focuses on the coordinated energy bidding of EH
in hybrid trading markets, and robust scheduling strategies hedging against worst-case uncertain distributions in different timescales,
the multi-energy management model without considering nonlinear conversion features \cite{zhao2018strategic}, \cite{zhu2020interval}
and startup shutdown costs \cite{al2021two}, \cite{8897020} can also be expressed.
Interested readers can refer to \cite{yong2021day} and \cite{ye2020incorporating} for more information about needed solution methods.

In addition, trade capacity constraints are also considered as follows:
\begin{eqnarray}
&&{{p}_{ie,k}}E_{in}^{\min }\le {{m}_{e,k}}\le {{p}_{ie,k}}E_{in}^{\max },\
\label{eq12}\\
&&{{p}_{ig,k}}G_{in}^{\min }\le {{m}_{g,k}}\le {{p}_{ig,k}}G_{in}^{\max },\
\label{eq13}\\
&&-E_{T,k}^{\max }\le {{E}_{T,k}}\le E_{T,k}^{\max },\
\label{eq14}
\end{eqnarray}
where $E_{in}^{\max}$ $(G_{in}^{\max})$, $E_{in}^{\min}$ $(G_{in}^{\min})$ denote upper and lower bounds for the electricity (natural gas) that can be bought from utility companies in day-ahead markets. $E_{T,k}^{\max}$ is the upper limit value of $E_{T,k}$ that is the electricity transacted in real-time 
power markets from other EH retailers with renewables.

To capture inner charge and discharge cycles of the power storage (denoted by $\psi=E$) and heat storage (denoted by $\psi=H$), we employ two kinds of positive variables $\psi_{s,k}^{ch}$, $\psi_{s,k}^{dis}$:
\begin{equation}
\psi_{s,k}=(\psi_{s,k}^{ch}-\psi_{s,k}^{dis})\Delta k, \psi \in\{E, H\}.
\label{eq15}
\end{equation}
The evolution constraints of energy storage state $S_{\psi,k}$ $(\psi \in\{E, H\})$ are shown as
\begin{eqnarray}
&&{{S}_{\psi,k}}={{S}_{\psi,k-1}}+\eta _{\psi}^{ch}\psi_{s,k}^{ch}\Delta k-\tfrac{1}{\eta _{\psi}^{dis}}\psi_{s,k}^{dis}\Delta k, \
\label{eq16}\\
&&0\le \psi_{s,k}^{ch}\le \psi_{s,k}^{ch,\max}, \
\label{eq17}\\
&&0\le \psi_{s,k}^{dis}\le \psi_{s,k}^{dis,\max}, \
\label{eq18}\\
&&S_{\psi,k}^{\min }\le {{S}_{\psi,k}}\le S_{\psi,k}^{\max}, \
\label{eq19}\\
&&{{S}_{\psi,T}}={{S}_{\psi,0}}, \
\label{eq20}
\end{eqnarray}
where $\eta _{\psi}^{ch}\!\in\!(0,1)$ ($\eta _{\psi}^{dis}\!\in\!(0,1)$) and $\psi_{s,k}^{ch,\max}$ ($\psi_{s,k}^{dis,\max}$) are the charge (discharge) efficiencies and maximum charge (discharge) rates of storage $\psi$;
$S_{\psi,k}^{\min}$ and $S_{\psi,k}^{\min}$ are the lower and upper limits of the energy storage level.
The energy state of storage $\psi$ in the final slot $S_{\psi,T}$ is set to the same value of the initial storage level $S_{\psi,0}$, which can also be other feasible states.
Note that storages are not allowed to be charged and discharged simultaneously due to physical workings, i.e., $\psi_{s,k}^{ch}\psi_{s,k}^{dis}\!=\!0$. But according to \cite{7792609,8731737}, this constraint would be redundant for the system optimization when $\eta _{\psi}^{ch}\eta _{\psi}^{dis}\!<\!1$.
\section{Day-ahead Level Energy Management Mechanism}\label{sec24}
This section provides a detailed formulation of day-ahead energy management under uncertainties in real-time power prices, renewable outputs and loads. First of all, mathematical models of uncertainties are introduced. Taking into account ambiguous PDFs of uncertainties, a two-stage optimization is developed to minimize the expected cost of EH by participating in day-ahead power/gas/carbon markets and real-time power markets and scheduling energy utilization. In the first stage, EH bids the quantity of needed electricity, gas and carbon credit to the day-ahead energy markets while the uncertainties have not been realized. In the second stage, it makes referential dispatch decisions to optimize the operation cost of energy sources within EH and the transaction profit (cost) in real-time power markets against the worst-case realization of uncertain distributions.
\subsection{Characterization of Uncertainties}
It is pointed out that this work mainly focuses on uncertain parameters about the electricity price of real-time market $p_{c,k}$, renewable generation $u_{f,k}$ and inelastic power load $D_{e,k}^{u}$ in the day-ahead schedule which are resulted from frequent dynamic changes and complex meteorological environment. Generally, day-ahead electricity and natural gas prices are considered as definite values during time series because the fluctuation of day-ahead market is ordinarily less compelling and could be effectively estimated~\cite{zhu2020interval}. Likewise, characterized by the preferable regularity and slow change, the inelastic heat load in EH $D_{h,k}^{u}$ is supposed to be a known quantity.
Under such circumstances, users' electricity needs are ensured to be satisfied with a certain level of confidence. In this way, the chance constraint is employed to formulate the electricity supply, i.e., $\mathbb{P}_{k,c}\{D_{e,k}+E_{T,k}-E_{out,k}\leq 0\}\geq 1-\varepsilon$. Subsequently, substituting \eqref{eq1}, \eqref{eq2} and \eqref{eq5}-\eqref{eq7} into the previously proposed chance constraint, we have
\begin{equation}
\mathbb{P}_{k,c}\{-{{\eta }_{T}}{{u}_{f,k}}+D_{e,k}^{u}+D_{e,k}^{s}+{{E}_{T,k}}-{{\eta }_{T}}\tfrac{{{m}_{e,k}}}{{{p}_{ie,k}}}-\tfrac{{{\eta }_{1}}{{\eta }_{G}}}{{{p}_{ig,k}}}{{m}_{g,k}}+{{\eta }_{1}}{{H}_{s,k}}+{{E}_{s,k}}+{{\eta }_{1}}D_{h,k}^{u}+{{\eta }_{1}}D_{h,k}^{s}\le 0\}\ge 1-\varepsilon,
\label{eq21}
\end{equation}
where $\eta_1=\eta_{M}^{e}/(\eta_G-\eta_{M}^{h})$; $1\!-\!\varepsilon$ is the confidence level of the chance constraint with $\varepsilon\!\in\!(0,1)$; $\mathbb{P}_{k,c}$ is the probability distribution of uncertainties which is introduced in the subsequent section.

As for the price uncertainty in real-time markets (denoted as ${{\xi }_{k,o}}={{p}_{c,k}}$), it is just embodied in the objective function and impacts the decision-performance optimality rather than the operational feasibility of EH.
In contrast, the uncertain renewable output and inelastic power load (denoted as ${{\bm{\xi }}_{k,c}}\!=\!{{[{{u}_{f,k}}\ D_{e,k}^{u}]}^{\rm T}}$ for simplicity) appearing in physical operation constraints exert noticeable influences on both the optimality of the solution and its feasibility in practice.
Moreover, the correlation of uncertain real-time power price with local small-size renewable generation is negligible when compared to the whole large-scale bidding pool according to \cite{7307233}. Besides, the inelastic power load is usually non-price stimulation so that the association between it and real-time power price can be ignored.
Consequently, we can separately model these uncertainties with the underlying probability distributions by introducing the following ambiguity sets ${\mathcal{P}}_{k,o}$ and ${\mathcal{P}}_{k,c}$ \cite{shang2018distributionally}:
\begin{eqnarray}
&&{{\mathcal{P}}_{k,o}}=\left\{ {{\mathbb{P}}_{k,o}}\in {{\mathfrak{B}}_{o}}\left| \begin{aligned}
  & \int_{{{\xi }_{k,o}}\in {{\mathbb{R}}^{L,o}}}{f({{\xi }_{k,o}})d{{\xi }_{k,o}}=1},\quad f({{\xi }_{k,o}})\ge 0 \\
 & {{\left[ {{\mathbb{E}}_{{{\mathbb{P}}_{k,o}}}}({{\xi }_{k,o}})-{{u}_{k,o}} \right]}^{2}}\Sigma _{k,o}^{-1}\le {{\gamma }_{k,o1}} \\
 & {{\mathbb{E}}_{{{\mathbb{P}}_{k,o}}}}[{{({{\xi }_{k,o}}-{{u}_{k,o}})}^{2}}]\le {{\gamma}_{k,o2}}{{\Sigma }_{k,o}} \\
\end{aligned} \right. \right\},\
\label{eq22}\\
&&{{\mathcal{P}}_{k,c}}=\left\{ {{\mathbb{P}}_{k,c}}\in {{\mathfrak{B}}_{c}}\left| \begin{aligned}
  & \int_{{{\bm{\xi }}_{k,c}}\in {{\mathbb{R}}^{L,c}}}{f({{\bm{\xi }}_{k,c}})d{{\bm{\xi }}_{k,c}}=1},\quad f({{\bm{\xi }}_{k,c}})\ge 0 \\
 & {{\left[ {{\mathbb{E}}_{{{\mathbb{P}}_{k,c}}}}({{\bm{\xi }}_{k,c}})-{{\bm{u}}_{k,c}} \right]}^{T}}\bm{\Sigma}_{k,c}^{-1}\left[ {{\mathbb{E}}_{{{\mathbb{P}}_{k,c}}}}({{\bm{\xi }}_{k,c}})-{{\bm{u}}_{k,c}} \right]\le {{\gamma }_{k,c1}} \\
 & {{\mathbb{E}}_{{{\mathbb{P}}_{k,c}}}}[({{\bm{\xi }}_{k,c}}-{{\bm{u}}_{k,c}}){{({{\bm{\xi }}_{k,c}}-{{\bm{u}}_{k,c}})}^{T}}]\le {{\gamma }_{k,c2}}{{\bm{\Sigma}}_{k,c}} \\
\end{aligned} \right. \right\},\
\label{eq23}
\end{eqnarray}
where $f({{\xi }_{k,o}})$ $(f({{\bm{\xi }}_{k,c}}))$, ${{\mathfrak{B}}_{o}}$ $({{\mathfrak{B}}_{c}})$ and ${{\mathbb{R}}^{L,o}}$ $({{\mathbb{R}}^{L,c}})$ are the probability density functions, nonnegative measures sets \cite{van2016generalized} and supports \cite{shang2018distributionally} of uncertainties ${{\xi }_{k,o}}$ $({{\bm{\xi }}_{k,c}})$ respectively;
${{u}_{k,o}}$ and ${{\Sigma}_{k,o}}$ (${{\bm{u}}_{k,c}}$ and ${{\bm{\Sigma}}_{k,c}}$) are the estimated average and variance values (the mean vector and covariance matrix) of ${{\xi }_{k,o}}$ (${{\bm{\xi }}_{k,c}}$) from historical samples. In fact, their accurate true values with the underlying probability distributions are difficult to obtain. To address this moment information error,
the true averages (mean vectors)
are assumed to be within ellipsoids centered at the estimated ${{u}_{k,o}}$ (${{\bm{u}}_{k,c}}$) and bounded by ${{\gamma }_{k,o1}}\!\geq\!0$ $({{\gamma }_{k,c1}}\!\geq\!0)$.
And true second moments (variances or covariance matrices)
are supposed to lie in positive semi-definite cones bounded by ${{\gamma}_{k,o2}}{{\Sigma }_{k,o}}$ $({{\gamma }_{k,c2}}{{\bm{\Sigma}}_{k,c}})$, with ${{\gamma}_{k,o2}}\!\geq\!\max\{{\gamma}_{k,o1},1\}$ $({{\gamma }_{k,c2}}\!\geq\!\max\{{\gamma}_{k,c1},1\})$.
Specifically, ${\gamma}_{k,o1}, {\gamma}_{k,o2}$ $({\gamma}_{k,c1}, {\gamma}_{k,c2})$ reflect the extent of closeness to the estimated mean and variance.
In other words, the bigger these parameters are the more conservative/robust the system may be.
A data-driven parameter selection method has been discussed in \cite{Erick2010} according to the size of samples, confidence region, risk level.
\cite{Erick2010} demonstrates that when the sample size goes to infinite the empirical values converge to the ones under true probability distributions.
Considering the above factors, we can first divide the sample into two subsets, and obtain parameter values and the ambiguity set for one of the subsets based on \cite{Erick2010}. Then we verify whether the second subset satisfies the ambiguity set by using its own estimated moment information.
Repeat and select appropriate  parameters step-by-step until the second subset belongs to the constructed ambiguity set.
\subsection{Two-stage Day-ahead Optimization}
According to aforementioned descriptions at the beginning of Section \ref{sec24},
it is noted that the energy purchase strategy ${{\bm{x}}_{1,k}}\!=\!\{{{m}_{e,k}},{{m}_{g,k}},{{m}_{c,k}}\}$ in day-ahead markets should be done at the first stage before uncertain parameters are revealed.
Besides, the day-ahead bidding strategy influences the optimal intra-day operation schedule of the EH, i.e., the second-stage decision $\bm{x}_{2,k}\!=\!\{E_{s,k},H_{s,k},D_{e,k}^{s},D_{h,k}^{s},E_{T,k}\}$ related to real-time markets, considering the uncertainties realizations in worst-case probability distributions.
Hence, we formulate the following DRO problem to incorporate stochastic supply-demand and intra-day transaction risks with hedging the distributional ambiguity.
\begin{eqnarray}
&&\underset{{{\bm{x}}_{1,k}}}{\mathop{\min }}\,\sum\limits_{k=1}^{T}{\left\{ \left( {{m}_{e,k}}+{{m}_{g,k}}+{{m}_{c,k}}+{{f}_{c,k}} \right)+\underset{f({{\xi }_{k,o}})\in {{\mathcal{P}}_{k,o}}}{\mathop{\max}}\,\int\limits_{{{\xi }_{k,o}}\in {{\mathbb{R}}^{L,o}}}{\underset{{{\bm{x}}_{2,k}}}{\mathop{\min }}\,\left( {{f}_{s,k}}-{{f}_{d,k}}-{{f}_{m,k}} \right)f({{\xi }_{k,o}})d{{\xi }_{k,o}}} \right\}}\
\notag\\
&&{\rm s.t.}\quad\eqref{eq1}-\eqref{eq23},\  \forall k,
\label{eq24}\\
&&\quad\quad{{f}_{s,k}}=a_{s,k}^{E}E_{s,k}^{2}+a_{s,k}^{H}H_{s,k}^{2},\
\label{eq25}\\
&&\quad\quad{{f}_{d,k}}={{\alpha }_{e,k}}{{(D_{e,k}^{s})}^{2}}+{{\beta }_{e,k}}D_{e,k}^{s}+{{\alpha }_{h,k}}{{(D_{h,k}^{s})}^{2}}+{{\beta }_{h,k}}D_{h,k}^{s},\
\label{eq26}\\
&&\quad\quad{{f}_{m,k}}={{p}_{c,k}}{{E}_{T,k}},\
\label{eq27}\\
&&\quad\ \,\,{{f}_{c,k}}\!=\!\left\{ \begin{aligned}
  &{{K}_{c,k}}(\tfrac{C_{o2e}}{p_{ie,k}}{m_{e,k}}\!+\!\tfrac{C_{o2g}}{p_{ig,k}}{{m}_{g,k}}\!-\!{{\eta }_{k}}{{P}_{lf,k}})\!-\!{{m}_{c,k}};\quad \tfrac{C_{o2e}}{p_{ie,k}}{{m}_{e,k}}\!+\!\tfrac{C_{o2g}}{p_{ig,k}}{m_{g,k}}\!\le\! \tfrac{{m}_{c,k}}{K_{c,k}}\!+\!{{\eta }_{k}}{{P}_{lf,k}} \\
 &{{K}_{H,k}}(\tfrac{C_{o2e}}{p_{ie,k}}{m_{e,k}}\!+\!\tfrac{C_{o2g}}{p_{ig,k}}{m_{g,k}}\!-\!{{\eta }_{k}}{{P}_{lf,k}}\!-\!\tfrac{m_{c,k}}{K_{c,k}});\ \;\, \tfrac{C_{o2e}}{p_{ie,k}}{m_{e,k}}\!+\!\tfrac{C_{o2g}}{p_{ig,k}}{{m}_{g,k}}\!\ge\! \tfrac{m_{c,k}}{K_{c,k}}\!+\!{{\eta}_{k}}{{P}_{lf,k}} \\
\end{aligned} \right.\
\label{eq28}
\end{eqnarray}
where formulas \eqref{eq25}-\eqref{eq28} represent the corresponding cost (revenue) terms of the objective function.
In \eqref{eq25}, the operation (degradation) cost of power and heat storages $f_{s,k}$ is modeled by the quadratic function with amortized coefficients $a_{s,k}^{E}$ and $a_{s,k}^{H}$.
$f_{d,k}$ in \eqref{eq26} is the utilities of elastic power and heat loads with positive linear factors ${{\beta}_{e,k}}$ and ${{\beta}_{h,k}}$ and negative quadratic factors ${{\alpha}_{e,k}}$ and ${{\alpha}_{h,k}}$. $f_{m,k}$ in \eqref{eq27} is the revenue obtained from the energy transaction (i.e., sale or buy) in the intra-day power market with the real-time price $p_{c,k}$.

Especially, $m_{c,k}$ plus $f_{c,k}$ in \eqref{eq28}
denotes the carbon trading cost taking into account the emission credits purchasing in carbon markets and fines on excessive emissions of carbon dioxide.
As for carbon markets, one of the fundamental mechanisms is the government-involved cap-and-trade system which contains the initial allocation of emission rights, carbon emission credits trading and settlement.
First, the government evaluates the total carbon emissions in a region, and divides them into several parts, known as a carbon emission credit which is freely allocated to market players based on relevant policy rules.
In this work, EH possesses some initial emission credits $E_{q,k}$
assigned by the local government,
i.e., $E_{q,k}\!=\!\eta_{k}P_{lf,k}$ where $\eta_{k}$ is the emission quota of unit-energy; $P_{lf,k}$ is the historical estimated total energy demands.
After that, the carbon emission credit allocation could be traded in carbon markets.
Let $E_{H,k}$ denote the extra emission credits purchased in the carbon market, and satisfy $E_{H,k}\!=\!u_{c,k}E_{q,k}$ where $u_{c,k}$ is the purchase margin. Then we formulate the corresponding expense as $m_{c,k}\!=\!K_{c,k}E_{H,k}$ with $K_{c,k}$ the carbon trading price in day-ahead markets.
In order to encourage the development of renewable energy, the carbon emissions $E_{p,k}$ of EH are supposed to be proportional to the use of coal-fired power with carbon intensity ${C_{o2e}}$ and natural gas with carbon intensity ${C_{o2g}}$, i.e., $E_{p,k}\!=\!{C_{o2e}}\tfrac{m_{e,k}}{p_{ie,k}}\!+\!{C_{o2g}}\tfrac{m_{g,k}}{p_{ig,k}}$.
Specifically, there are three scenarios in trade settlement. Firstly, if the carbon emission is less than initial carbon credits ($E_{p,k}\!\leq\!E_{q,k}$), the superfluous credits can be sold to obtain revenues, and the corresponding carbon cost is ${{K}_{c,k}}\!\cdot\![{{E}_{p,k}}\!-\!({{E}_{H,k}}\!+\!{{E}_{q,k}})]\!+\!m_{c,k}$.
Secondly, if the carbon emission is more than initial carbon credits and less than the purchased part (${{E}_{q,k}}\!\le\!{{E}_{p,k}}\!\le\!{{E}_{H,k}}\!+\!{{E}_{q,k}}$), EH only pays for the part that runs out of the initially allocated credits. And the corresponding carbon cost is ${{K}_{c,k}}\!\cdot\![{{E}_{p,k}}\!-\!({{E}_{H,k}}\!+\!{{E}_{q,k}})]\!+\!m_{c,k}$.
Otherwise, if carbon emissions exceed the total holdings of carbon credits (${{E}_{p,k}}\!\ge\! {{E}_{H,k}}\!+\!{{E}_{q,k}}$), EH needs to pay for not only the purchased emission rights but also the penalty of over-emission with the unit price $K_{H,k}$. In this case, the corresponding carbon cost is ${{K}_{H,k}}\!\cdot\![{{E}_{p,k}}\!-\!({{E}_{H,k}}\!+\!{{E}_{q,k}})]\!+\!m_{c,k}$. To sum up, combining with the aforementioned terms,
total carbon costs are rearranged into $m_{c,k}+f_{c,k}$ as shown in \eqref{eq24} and \eqref{eq28}.
\section{Intra-day Energy Management Mechanism}\label{sec25}
\begin{figure}[htbp]
\centering
\includegraphics[width=4.39in]{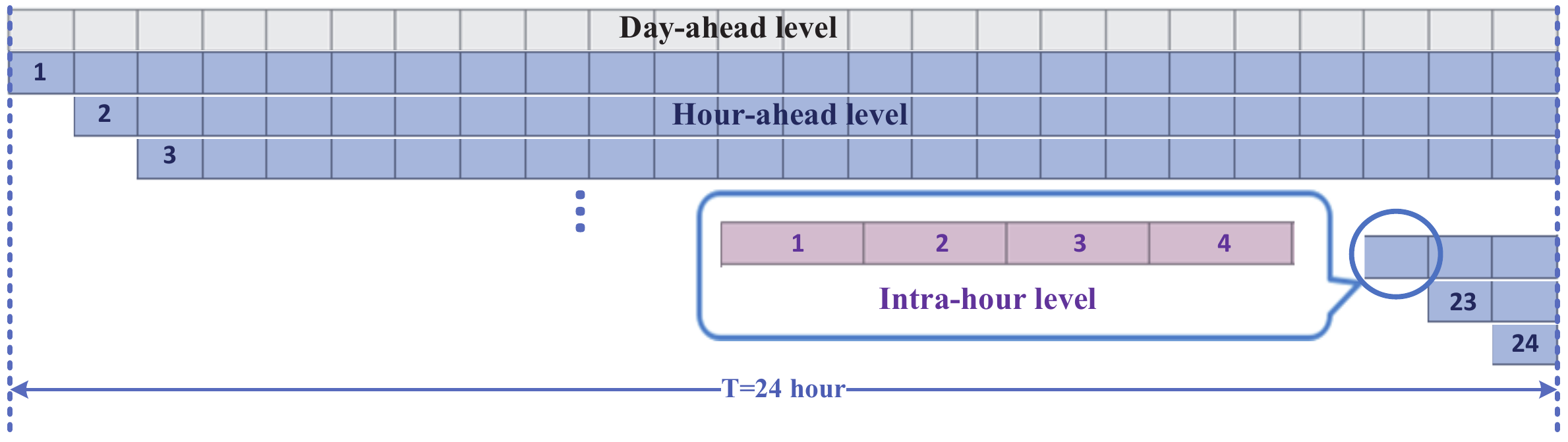}
\caption{Schematic time length of the proposed three-level optimization.}
\label{fig2}
\end{figure}

Once day-ahead scheduling results are obtained, EH implements the bidding strategy \begin{small}{${{\bm{x}}_{1}}$}\end{small} in day-ahead markets and makes the energy dispatch strategy \begin{small}{${{\bm{x}}_{2}}$}\end{small} as a reference for intra-day dispatch.
Due to distribution uncertainties and fluctuations of renewable generation, power demand and real-time market prices, the dispatch strategy in day-ahead schedule may diverge from the optimal one.
Additionally, the energy dispatch regulation of intra-day usually refers to multiple timescales owing to different response time of energy components.
For example, the intrinsic thermal inertia and settling process of heat load, heat storage and heat conversion equipment facilitate a slow timescale dispatch (e.g., one hour).
Besides, time-varying real-time electricity prices are released hourly which motivates the intra-day power transaction to be established hour-ahead.
Contrastively, the operations of power storage and electricity load run in relatively rapid processes.
On one hand, dispatching all energy components in a single slow timescale may result in less accurate electricity dispatches.
On the other hand, the solution dimension becomes too high if a single rapid timescale is adopted.
Against this background, based on the scheduling results of day-ahead level, we move on to re-optimize the intra-day energy dispatch respectively under two timescales (i.e., hour-ahead level with a one-hour time step and intra-hour level with a 15-minute time step) employing updated data related to uncertainties. The time length of aforementioned three-level optimization period is illustrated in Fig.\ref{fig2}.
\subsection{Hour-ahead Level Optimization}
In the intra-day, real-time electricity market prices are settled in sequence and ahead of the trading hour $\tau$ as time goes on. Moreover, the prediction moment information and corresponding ambiguity sets of renewable generation, power demand ($k\!\geq\!\tau$) and real-time electricity prices in the remaining time ($k\!>\!\tau$) are updated (e.g., based on extended short-term forecasting technique).
To revise the day-ahead energy dispatch strategies, hour-ahead problem re-optimizes the scheduling from the following hour $\tau$ to the future end time $T$ based on the system operation status and the updated information of uncertainties.
\begin{equation}
\begin{aligned}
&\underset{\{\bm{x}_{ha,k}\}}{\mathop{\min}}\Big\{\sum\nolimits_{k=\tau}^{k=\tau}(f_{s,k}-f_{d,k}+f_{s,k}^{pn}+f_{d,k}^{pn}-f_{m,k})+\\
&\phantom{=\;\;}\sum\nolimits_{k>\tau}^{T}{\underset{f(\tilde{\xi}_{k,o})\in{\mathcal{P}_{k,o}}}{\mathop{\max}}}\int\limits_{\tilde{\xi}_{k,o}\in{\mathbb{R}}^{L,o}}\underset{\{\bm{x}_{ha,k}\}}{\mathop{\min}}(f_{s,k}-f_{d,k}+f_{s,k}^{pn}+f_{d,k}^{pn}-f_{m,k})f(\tilde{\xi}_{k,o})d{\tilde{\xi}_{k,o}}\Big\},
\end{aligned}
\label{eq29}
\end{equation}
where decision variables ${{\bm{x}}_{ha,k}}=\{{{\tilde{E}}_{s,k}},{{\tilde{H}}_{s,k}},\tilde{D}_{e,k}^{s},\tilde{D}_{h,k}^{s},{{\tilde{E}}_{T,k}}\}$ 
are similar to ${{\bm{x}}_{2,k}}$
in the day-ahead problem with tilde symbols representing the hour-ahead level. The storage operation cost ${f}_{s,k}$, demand utility ${f}_{d,k}$ and intra-day transaction revenue ${f}_{m,k}$ are defined in the same manner as day-ahead optimal functions.
$f_{s,k}^{pn}$ and $f_{d,k}^{pn}$ are penalty costs of deviations between the storage and demand schedules respectively, which associates the day-ahead schedules with hour-ahead level managements, in favor of the healthy system operation.
These costs are expressed as follows with corresponding penalty coefficients $C_{E}^{pn}$, $C_{H}^{pn}$, $C_{D_{e}}^{pn}$, $C_{D_{h}}^{pn}$\footnote{Note that the unit of these penalty coefficients is \textcent/$(kW)^{2}$, and the time slot $\Delta k=1$h.};
\begin{eqnarray}
&&f_{s,k}^{pn}=C_{E}^{pn}({\tilde{E}}_{s,k}/\Delta k-E_{s,k}/\Delta k)^{2}+C_{H}^{pn}({\tilde{H}}_{s,k}/\Delta k-H_{s,k}/\Delta k)^{2},\
\label{eq30}\\
&&f_{d,k}^{pn}=C_{D_{e}}^{pn}(\tilde{D}_{e,k}^{s}/\Delta k-D_{e,k}^{s}/\Delta k)^{2}+C_{D_{h}}^{pn}(\tilde{D}_{h,k}^{s}/\Delta k-D_{h,k}^{s}/\Delta k)^{2}.\
\label{eq31}
\end{eqnarray}

Besides, operation constraints of hour-ahead level ($k\geq\tau$) are expressed as follows:
\begin{eqnarray}
&&\left\{ \begin{aligned}
&{{p}_{ig,k}}(D_{h,k}^{u}+\tilde{D}_{h,k}^{s}+{{\tilde{H}}_{s,k}})-{{\eta }_{G}}{{m}_{g,k}}\le 0, \\
&{{m}_{g,k}}\eta _{M}^{h}-{{p}_{ig,k}}(D_{h,k}^{u}+\tilde{D}_{h,k}^{s}+{{\tilde{H}}_{s,k}})\le 0, \\
&{{\eta }_{G}}{{m}_{g,k}}-{{p}_{ig,k}}(D_{h,k}^{u}+\tilde{D}_{h,k}^{s}+{{\tilde{H}}_{s,k}})-G_{MT}^{\max }{{p}_{ig,k}}({{\eta }_{G}}-\eta _{M}^{h})\le 0, \\
&{{p}_{ig,k}}(D_{h,k}^{u}+\tilde{D}_{h,k}^{s}+{{\tilde{H}}_{s,k}})-{{m}_{g,k}}\eta _{M}^{h}-G_{GF}^{\max }{{p}_{ig,k}}({{\eta }_{G}}-\eta _{M}^{h})\le 0, \\
\end{aligned} \right.\
\label{eq32}\\
&&-E_{T,k}^{\max}\leq{{\tilde{E}}_{T,k}}\leq E_{T,k}^{\max}, \
\label{eq33}\\
&&\tilde{H}_{s,k}=(\tilde{H}_{s,k}^{ch}-\tilde{H}_{s,k}^{dis})\Delta k,  \
\label{eq34}\\
&&\tilde{S}_{H,k}=\tilde{S}_{H,k-1}+\eta_{H}^{ch}\tilde{H}_{s,k}^{ch}\Delta k-\tfrac{1}{\eta _{H}^{dis}}\tilde{H}_{s,k}^{dis}\Delta k, \
\label{eq35}\\
&&0\le \tilde{H}_{s,k}^{ch}\le H_{s,k}^{ch,\max}, \, 0\le \tilde{H}_{s,k}^{dis}\le H_{s,k}^{dis,\max}, \
\label{eq36}\\
&&S_{H,k}^{\min }\le \tilde{S}_{H,k}\le S_{H,k}^{\max}, \, \quad\tilde{S}_{H,T}=\tilde{S}_{H,0},\
\label{eq37}\\
&&\tilde{E}_{s,k}=(\tilde{E}_{s,k}^{ch}-\tilde{E}_{s,k}^{dis})\Delta k,  \,
\label{eq38}\\
&&\tilde{S}_{E,k}=\tilde{S}_{E,k-1}+\eta_{E}^{ch}\tilde{E}_{s,k}^{ch}\Delta k-\tfrac{1}{\eta _{E}^{dis}}\tilde{E}_{s,k}^{dis}\Delta k, \
\label{eq39}\\
&&0\le \tilde{E}_{s,k}^{ch}\le E_{s,k}^{ch,\max}, \, 0\le \tilde{E}_{s,k}^{dis}\le E_{s,k}^{dis,\max}, \
\label{eq40}\\
&&S_{E,k}^{\min }\le \tilde{S}_{E,k}\le S_{E,k}^{\max}, \quad \tilde{S}_{E,T}=\tilde{S}_{E,0},\quad \tilde{S}_{E,\tau-1}=\tilde{\tilde{S}}_{E,\tau-1,4},\
\label{eq41}\\
&&\sum\nolimits_{k\ge\tau}^{T}{\tilde{D}_{e,k}^{s}}\!\ge\! D_{e}^{s,day}\!-\!\sum\nolimits_{k<\tau}\sum\nolimits_{m=1}^{4}{\tilde{\tilde{D}}_{e,k,m}^{s}},\,\; \
\label{eq42}\\
&&D_{e,k}^{s,\min}\Delta k\!\leq\! \tilde{D}_{e,k}^{s}\!\leq\! D_{e,k}^{s,\max}\Delta k,\,\; \!-\!D_{e}^{r}\Delta k\!\leq\! \tilde{D}_{e,k+1}^{s}\!-\!\tilde{D}_{e,k}^{s}\!\leq\! D_{e}^{r}\Delta k,\, \
\label{eq43}\\
&&\sum\nolimits_{k\ge\tau}^{T}{\tilde{D}_{h,k}^{s}}\!\ge\! D_{h}^{s,day}\!-\!\sum\nolimits_{k<\tau}{\tilde{D}_{h,k}^{s}},\,\
\label{eq44}\\
&&D_{h,k}^{s,\min}\Delta k\!\leq\! \tilde{D}_{h,k}^{s}\!\leq\! D_{h,k}^{s,\max}\Delta k,\,\; \!-\!D_{h}^{r}\Delta k\!\leq\! \tilde{D}_{h,k+1}^{s}\!-\!\tilde{D}_{h,k}^{s}\!\leq\! D_{h}^{r}\Delta k,\,\
\label{eq45}\\
&&{\begin{aligned}\mathbb{P}_{k,c}\big\{-&{{\eta }_{T}}{\tilde{u}_{f,k}}+\tilde{D}_{e,k}^{u}+\tilde{D}_{e,k}^{s}+\tilde{E}_{T,k}-\tfrac{{{\eta }_{T}}}{{{p}_{ie,k}}}{m_{e,k}}-\tfrac{{{\eta }_{1}}{{\eta }_{G}}}{{{p}_{ig,k}}}{{m}_{g,k}}+\\
&{{\eta }_{1}}{\tilde{H}_{s,k}}+{\tilde{E}_{s,k}}+{{\eta}_{1}}D_{h,k}^{u}+{{\eta }_{1}}\tilde{D}_{h,k}^{s}\le 0\big\}\ge 1-\varepsilon,\;\, \
\end{aligned}}
\label{eq46}
\end{eqnarray}
where \eqref{eq41} and \eqref{eq42} deal with the multi-period storage and elastic demand models and ensure state continuity for the cyclic dispatch between two timescales.
$\tilde{\tilde{S}}_{E,\tau-1,4}$ of \eqref{eq41} denotes the practical initial power storage state of the present time, i.e., the optimal power storage results on the corresponding intra-hour level.
In \eqref{eq42}, ${\tilde{\tilde{D}}_{e,k,m}^{s}}$ is the actual implemented elastic power demand during previous time slots which is also equivalent to the optimal value of intra-hour level.
The rest hour-ahead operation constraints \eqref{eq32}-\eqref{eq40} and \eqref{eq43}-\eqref{eq46} are similar to the constraints in day-ahead problems with the additional symbol `$\sim$' denoting the adjusted decision variables in hour-ahead level under updated uncertainty information and current system status.

It should be pointed out that, at the hour-ahead level, only decisions for the next time slot $k\!=\!\tau$ will be preserved, and just with $\{\tilde{H}_{s,k},\tilde{D}_{h,k}^{s},\tilde{E}_{T,k}\}$ implemented in practice. Others $k\!>\!\tau$ will be deserted. Repeating the above processes with a one-hour time step, the system can be re-optimized in a receding way to revise day-ahead energy scheduling strategies.
\subsection{Intra-hour Level Optimization}
At the intra-hour level with a 15-minute timescale, the objective is to address operation discrepancy and unbalanced state induced by the fluctuation (variation) of renewable generation (inelastic power demand).
Finally, the actual energy management of elastic power load and electricity storage is settled based on the current system status and the corresponding optimal results of hour-ahead and day-ahead levels. The 
optimization model is shown in the following\footnote{Different time steps are considered in hour-ahead and intra-hour levels: $\Delta k=1$h, $\Delta m=\tfrac{1}{4}$h.}:
\begin{eqnarray}
&&{\begin{aligned}\underset{{{\bm{x}}_{ih,k,m}}}{\mathop{\min }}\,&C_{D_{e}}^{pn}(\tilde{\tilde{D}}_{e,k,m}^{s}/\Delta m-\tilde{D}_{e,k}^{s}/\Delta k)^{2}-[\alpha_{e,k}(\tilde{\tilde{D}}_{e,k,m}^{s})^{2}+\beta_{e,k}\tilde{\tilde{D}}_{e,k,m}^{s}]+\\
&\qquad C_{E}^{pn}(\tilde{\tilde{E}}_{s,k,m}/\Delta m-\tilde{E}_{s,k}/\Delta k)^{2}+a_{s,k}^{E}(\tilde{\tilde{E}}_{s,k,m})^{2}\
\end{aligned}}
\label{eq47}\\
&&{\begin{aligned}{\rm s.t.}\;&-\eta_{T}\tilde{\tilde{u}}_{f,k,m}\tfrac{\Delta k}{\Delta m}+\tilde{\tilde{D}}_{e,k,m}^{u}\tfrac{\Delta k}{\Delta m}+\tilde{\tilde{D}}_{e,k,m}^{s}\tfrac{\Delta k}{\Delta m}+\tilde{E}_{T,k}-\tfrac{\eta_{T}}{p_{ie,k}}m_{e,k}-\tfrac{\eta_{1}\eta_{G}}{p_{ig,k}}m_{g,k}\\
&\qquad\quad+\eta_{1}\tilde{H}_{s,k}+\tilde{\tilde{E}}_{s,k,m}\tfrac{\Delta k}{\Delta m}+\eta_{1}D_{h,k}^{u}+\eta_{1}\tilde{D}_{h,k}^{s}=0,\
\end{aligned}}
\label{eq48}\\
&&\qquad\, D_{e,k}^{s,\min}\Delta m\!\leq\! \tilde{\tilde{D}}_{e,k,m}^{s}\!\leq\! D_{e,k}^{s,\max}\Delta m,\,\; \!-\!D_{e}^{r}\Delta m\!\leq\! \tilde{\tilde{D}}_{e,k,m-1}^{s}\!-\!\tilde{\tilde{D}}_{e,k,m}^{s}\!\leq\! D_{e}^{r}\Delta m,\
\label{eq49}\\
&&\qquad\, \tilde{\tilde{E}}_{s,k,m}\!=\!(\tilde{\tilde{E}}_{s,k,m}^{ch}\!-\!\tilde{\tilde{E}}_{s,k,m}^{dis})\Delta m,\,\,\tilde{\tilde{S}}_{E,k,m}\!=\!\tilde{\tilde{S}}_{E,k,m\!-\!1}\!+\!\eta_{E}^{ch}\tilde{\tilde{E}}_{s,k,m}^{ch}\Delta m\!-\!\tfrac{1}{\eta_{E}^{dis}}\tilde{\tilde{E}}_{s,k,m}^{dis}\Delta m,\
\label{eq50}\\
&&\qquad\, 0\le \tilde{\tilde{E}}_{s,k,m}^{ch}\le E_{s,k}^{ch,\max}, \, 0\le \tilde{\tilde{E}}_{s,k,m}^{dis}\le E_{s,k}^{dis,\max}, \, S_{E,k}^{\min }\le \tilde{\tilde{S}}_{E,k,m}\le S_{E,k}^{\max},\
\label{eq51}
\end{eqnarray}
where ${{\bm{x}}_{ih,k,m}}\!=\!\{\tilde{\tilde{D}}_{e,k,m}^{s},\tilde{\tilde{E}}_{s,k,m}\}$ with subscript $k,m$ denoting the $m$-th slot of hour $k$.
The objective function \eqref{eq47} includes deviation penalty costs and utilities of elastic power demand,  deviation penalty and operation costs of power storage, the definitions of which are similar to those in the hour-ahead level. \eqref{eq48} represents the new electricity balance constraint coordinating the updated prediction of renewable generation and inelastic power load in a short slot $\Delta m$ (regarded as the actual operation value). The meanings of rest constraints \eqref{eq49}-\eqref{eq51} are the same as those in hour-ahead level. Moreover, it is noted that, in adjacent hours, the storage state is imposed by $\tilde{\tilde{S}}_{E,k-1,4}=\tilde{\tilde{S}}_{E,k,0}$ to ensure the consistency of optimal operations.
\section{Solution Methodology}\label{sec3}
The proposed coordinated energy management models in previous sections manifest the features of
non-linear (e.g., to deal with conditional piecewise carbon trading cost function),
chance-constrained (e.g., to deal with distribution uncertainties in power supply and demand), two stages (e.g., to deal with second-stage decision in day-ahead schedules hedging all of the probable outcomes about probability distributions of uncertainties in ambiguity sets).
Thus, the developed problems are generally intractable and difficult to directly calculate.
In this case, we propose subsequent appropriate transformation processes to tackle this complex problem.
\subsection{Linearized Reconstruction of Carbon Trading Function}
One difficulty in solving the carbon trading cost ${f}_{c,k}$ \eqref{eq28} is the presence of the segmental structure and conditional functions.
To linearize the original form and obtain an explicit reformulation,
we first introduce a binary variable with the indicator function: ${{z}_{k}}=\mathds{1}\{( \tfrac{{{C}_{o2e}}}{{{p}_{ie,k}}}{{m}_{e,k}}+\tfrac{{{C}_{o2g}}}{{{p}_{ig,k}}}{{m}_{g,k}}-\tfrac{1}{{{K}_{c,k}}}{{m}_{c,k}}-{{\eta }_{k}}{{P}_{lf,k}})\le 0 \}$, which means ${{z}_{k}}=1$ when the condition $\tfrac{{{C}_{o2e}}}{{{p}_{ie,k}}}{{m}_{e,k}}+\tfrac{{{C}_{o2g}}}{{{p}_{ig,k}}}{{m}_{g,k}}\le \tfrac{1}{{{K}_{c,k}}}{{m}_{c,k}}+{{\eta }_{k}}{{P}_{lf,k}}$ is met; otherwise ${{z}_{k}}=0$.
Besides, the above-mentioned logic constraint can be converted to the form \eqref{eq52} employing Big-M techniques.
\begin{equation}
-{{z}_{k}}M\le \tfrac{{{C}_{o2e}}}{{{p}_{ie,k}}}{{m}_{e,k}}+\tfrac{{{C}_{o2g}}}{{{p}_{ig,k}}}{{m}_{g,k}}-\tfrac{1}{{{K}_{c,k}}}{{m}_{c,k}}-{{\eta }_{k}}{{P}_{lf,k}}\le (1-{{z}_{k}})M,\
\label{eq52}
\end{equation}
where $M$ denotes a sufficiently large constant value \cite{luo2000conic}. To be specific, the value of $M$ can be selected according to trial and error combining with the physical nature limits of decision variables $m_{e,k}$, $m_{g,k}$, $m_{c,k}$.
Secondly, the segmental cost function \eqref{eq28} is reformulated as
\begin{equation}
\begin{aligned}
{{f}_{c,k}}=&{{z}_{k}}[{{K}_{c,k}}(\tfrac{{{C}_{o2e}}}{{{p}_{ie,k}}}{{m}_{e,k}}+\tfrac{{{C}_{o2g}}}{{{p}_{ig,k}}}{{m}_{g,k}})-{{K}_{c,k}}{{\eta }_{k}}{{P}_{lf,k}}-{{m}_{c,k}}]+ \\
& (1\!-\!{{z}_{k}})[{{K}_{H,k}}(\tfrac{{{C}_{o2e}}}{{{p}_{ie,k}}}{{m}_{e,k}}\!+\!\tfrac{{{C}_{o2g}}}{{{p}_{ig,k}}}{{m}_{g,k}})\!-\!{{K}_{H,k}}{{\eta }_{k}}{{P}_{lf,k}}\!-\!\tfrac{{{K}_{H,k}}}{{{K}_{c,k}}}{{m}_{c,k}}]. \\
\end{aligned}
\label{eq53}
\end{equation}

Unfortunately, there are bilinear terms resulted from the multiplication of binary variable $z_{k}$ and continuous variables $m_{e,k}$, $m_{g,k}$, $m_{c,k}$ in the function \eqref{eq53}, which imposes computational challenges. To tackle this issue, we employ auxiliary variables and introduce additional constraints based on Big-M approaches as follows.
\begin{eqnarray}
&&{{\sigma}_{{{m}_{e,k}}}}={{z}_{k}}{{m}_{e,k}},\;{{\sigma}_{{{m}_{g,k}}}}={{z}_{k}}{{m}_{g,k}},\;{{\sigma}_{{{m}_{c,k}}}}={{z}_{k}}{{m}_{c,k}}\
\label{eq54}\\
&&\left\{\begin{aligned}
  & -M{{z}_{k}}\le {{\sigma }_{{{m}_{e,k}}}}\le M{{z}_{k}} \\
 & {{m}_{e,k}}-M(1-{{z}_{k}})\le {{\sigma }_{{{m}_{e,k}}}}\le {{m}_{e,k}}+M(1-{{z}_{k}}) \\
\end{aligned} \right.
\label{eq55}\\
&&\left\{\begin{aligned}
  & -M{{z}_{k}}\le {{\sigma }_{{{m}_{g,k}}}}\le M{{z}_{k}} \\
 & {{m}_{g,k}}-M(1-{{z}_{k}})\le {{\sigma }_{{{m}_{g,k}}}}\le {{m}_{g,k}}+M(1-{{z}_{k}}) \\
\end{aligned} \right.
\label{eq56}\\
&&\left\{ \begin{aligned}
  & -M{{z}_{k}}\le {{\sigma }_{{{m}_{c,k}}}}\le M{{z}_{k}} \\
 & {{m}_{c,k}}-M(1-{{z}_{k}})\le {{\sigma }_{{{m}_{c,k}}}}\le {{m}_{c,k}}+M(1-{{z}_{k}}) \\
\end{aligned} \right.
\label{eq57}
\end{eqnarray}

After that, the function \eqref{eq53} can be rewritten as
\begin{equation}
\begin{aligned}
  & {{f}_{c,k}}=({{K}_{c,k}}-{{K}_{H,k}})(\tfrac{{{C}_{o2e}}}{{{p}_{ie,k}}}{{\sigma }_{{{m}_{e,k}}}}+\tfrac{{{C}_{o2g}}}{{{p}_{ig,k}}}{{\sigma }_{{{m}_{g,k}}}}-{{\eta }_{k}}{{P}_{lf,k}}{{z}_{k}}) \\
 & \quad \quad +{{K}_{H,k}}(\tfrac{{{C}_{o2e}}}{{{p}_{ie,k}}}{{m}_{e,k}}\!+\!\tfrac{{{C}_{o2g}}}{{{p}_{ig,k}}}{{m}_{g,k}}\!-\!{{\eta }_{k}}{{P}_{lf,k}})\!+\!(\tfrac{{{K}_{H,k}}}{{{K}_{c,k}}}\!-\!1){{\sigma }_{{{m}_{c,k}}}}\!-\!\tfrac{{{K}_{H,k}}}{{{K}_{c,k}}}{{m}_{c,k}}. \\
\end{aligned}
\label{eq58}
\end{equation}

Eventually, the original complex segmental term is exactly linearized and transformed into a mixed-integer linear model which can be efficiently addressed.
\subsection{Transformation of Two-stage DRO}
The proposed distributionally robust two-stage model in day-ahead (hour-ahead) scheduling problems with the form of min-max-min immunized against the distributional uncertainty is untractable and cannot be resolved straightforwardly.
In order to tackle this issue, in the following, we take the day-ahead problem \eqref{eq24} as an example for brevity, and first rewrite the inner max-min term in \eqref{eq24} as the following explicit form:
\begin{eqnarray}
&&\underset{f({{\xi }_{k,o}})\in {{\mathcal{P}}_{k,o}}}{\mathop{\max}}\,\int_{{{\xi }_{k,o}}\in {{\mathbb{R}}^{L,o}}}{\Phi(\bm{x}_{2,k},{{\xi }_{k,o}})f({{\xi }_{k,o}})d{{\xi}_{k,o}}},\
\label{eq59}\\
&&\;\;\quad{\rm s.t.}\quad \int_{{{\xi }_{k,o}}\in {{\mathbb{R}}^{L,o}}}f({{\xi }_{k,o}})d{{\xi}_{k,o}}=1: r_{k},\
\label{eq60}\\
&&\;\;\quad\qquad\int_{{{\xi }_{k,o}}\in {{\mathbb{R}}^{L,o}}}({{\xi }_{k,o}}-u_{k,o})^{2}f({{\xi }_{k,o}})d{{\xi}_{k,o}}\leq {{\gamma}_{k,o2}}{{\Sigma }_{k,o}}: Q_{k},\
\label{eq61}\\
&&\quad\qquad\int_{{{\xi }_{k,o}}\in {{\mathbb{R}}^{L,o}}}\left[ \begin{matrix}
   {{\Sigma }_{k,o}} & {{\xi }_{k,o}}-u_{k,o}  \\
   {{\xi }_{k,o}}-u_{k,o} & {{\gamma}_{k,o1}}  \\
\end{matrix} \right]f({{\xi }_{k,o}})d{{\xi}_{k,o}}\geq 0: \left[ \begin{matrix}
   P_{k} & v_{k}  \\
   v_{k} & S_{k}  \\
\end{matrix} \right],\
\label{eq62}
\end{eqnarray}
where $\Phi(\bm{x}_{2,k},{{\xi }_{k,o}})\!=\!\min f_{s,k}\!-\!f_{d,k}\!-\!f_{m,k}\!=\!f_{s,k}\!-\!f_{d,k}\!-\!{{\xi }_{k,o}}E_{T,k}$,
The decision variable in the above infinite dimensional optimization problem is the continuous probability density function of uncertainty over ${{\mathbb{R}}^{L,o}}$.
Further, based on the theory of duality in cone programming \cite{shapiro2001duality,bertsimas2010models}, the above problem \eqref{eq59}-\eqref{eq62} can be reformulated as below \cite{delage2010distributionally} with corresponding dual variables shown at the right-hand side of constraints \eqref{eq60}-\eqref{eq62},
\begin{eqnarray}
&&\underset{r_{k},Q_{k},P_{k},v_{k},S_{k}}{\mathop{\min}}\,r_{k}+({{\gamma}_{k,o2}}{{\Sigma }_{k,o}}-u_{k,o}^{2})Q_{k}+({{\Sigma }_{k,o}}P_{k})-2u_{k,o}v_{k}+{{\gamma}_{k,o1}}S_{k},\
\label{eq63}\\
&&\qquad{\rm s.t.}\quad r_{k}+{{\xi }_{k,o}}^{2}Q_{k}-2{{\xi }_{k,o}}Q_{k}u_{k,o}-2{{\xi }_{k,o}}v_{k}-\Phi(\bm{x}_{2,k},{{\xi }_{k,o}})\geq0,\forall{{\xi }_{k,o}}\in {{\mathbb{R}}^{L,o}},\
\label{eq64}\\
&&\qquad\qquad Q_{k}\geq0,\
\label{eq65}\\
&&\qquad\qquad \left[ \begin{matrix}
   P_{k} & v_{k}  \\
   v_{k} & S_{k}  \\
\end{matrix} \right]\geq0.\
\label{eq66}
\end{eqnarray}

Next, we can fix variables $Q_{k}$, $r_{k}$ to analyze the optimal value of remainder variables $P_{k}$, $v_{k}$, $S_{k}$.
Note that the optimal value of $S_{k}$ meets $S_{k}^{*}\geq0$ as indicated in \eqref{eq66}.
Moreover, the results are discussed from two aspects.

1) When $S_{k}^{*}\!>\!0$, formula \eqref{eq66} is converted into $P_{k}\geq(1/S_{k})v_{k}^{2}$.
And the optimal value has $P_{k}^{*}\!=\!(1/S_{k})v_{k}^{2}$
due to ${{\Sigma }_{k,o}}\!\geq\!0$. Substituting this optimal solution into \eqref{eq63}, the objective is degraded to resolve the following term: $\min_{S_{k}>0}(1/S_{k})v_{k}^{2}{{\Sigma }_{k,o}}+{{\gamma}_{k,o1}}S_{k}$.
Besides, taking the first-order derivative of this term as zero, the optimal value of $S_{k}$ yields 
$S_{k}^{*}=\sqrt{(1/{{\gamma}_{k,o1}}){v_{k}^{2}{{\Sigma }_{k,o}}}}$.
Then by using the change of variable $w_{k}=-2(v_{k}+Q_{k}u_{k,o})$, the optimization function \eqref{eq63} is arranged as the following form
\begin{equation}
r_{k}+{{\gamma}_{k,o2}}{{\Sigma }_{k,o}}Q_{k}+u_{k,o}^{2}Q_{k}+u_{k,o}w_{k}+\sqrt{{\gamma}_{k,o1}}\left|{{\Sigma }_{k,o}^{1/2}}(w_{k}+2Q_{k}u_{k,o})\right|.\
\label{eq67}
\end{equation}

2) When $S_{k}^{*}\!=\!0$, the optimal value of $v_{k}$ meets $v_{k}^{*}\!=\!0$ which can be proved by reductio: if $v_{k}^{*}\!\neq\!0$, we have ${v_{k}^{*}}^{2}\!>\!0$
and $P_{k}^{*}S_{k}^{*}\!-\!{v_{k}^{*}}^{2}\!<\!0$
that contradicts with \eqref{eq66}. In this case, it is easy to verify $P_{k}^{*}\!=\!0$ as it minimizes \eqref{eq63} with coefficient ${{\Sigma }_{k,o}}\!\geq\!0$.
Again we use the change of variable $w_{k}\!=\!-2(v_{k}+Q_{k}u_{k,o})$ and conclude that when $S^{*}\!=\!0$, the objective of  \eqref{eq63} is also equivalent to the optimization problem 
$r_{k}+{{\gamma}_{k,o2}}{{\Sigma }_{k,o}}Q_{k}-u_{k,o}^{2}Q_{k}=r_{k}+{{\gamma}_{k,o2}}{{\Sigma }_{k,o}}Q_{k}+u_{k,o}^{2}Q_{k}+u_{k,o}w_{k}+\sqrt{{\gamma}_{k,o1}}\left|{{\Sigma }_{k,o}^{1/2}}(w_{k}+2Q_{k}u_{k,o})\right|$.

Specially, in view of the structure of formula $\Phi(\bm{x}_{2,k},{{\xi }_{k,o}})$ and using the variable $w_{k}$ in place of $-2(v_{k}+Q_{k}u_{k,o})$, the constraint \eqref{eq64} can be rewritten as
${{\xi}_{k,o}}^{2}Q_{k}+(w_{k}+E_{T,k}){{\xi }_{k,o}}+r_{k}-f_{s,k}+f_{d,k}\geq0$.
It is equivalent to the compact form as follows
\begin{equation}
\left[ \begin{matrix}
   {{\xi }_{k,o}}  \\
   1  \\
\end{matrix} \right]^{T}\left[ \begin{matrix}
   Q_{k} & \tfrac{w_{k}+E_{T,k}}{2}  \\
   \tfrac{w_{k}+E_{T,k}}{2} & r_{k}-f_{s,k}+f_{d,k}  \\
\end{matrix} \right]\left[ \begin{matrix}
   {{\xi }_{k,o}}  \\
   1  \\
\end{matrix} \right]\geq0, \forall{{\xi }_{k,o}}\in {{\mathbb{R}}^{L,o}}.\
\label{eq69}
\end{equation}
Thus we have
\begin{equation}
\left[ \begin{matrix}
   Q_{k} & \tfrac{w_{k}+E_{T,k}}{2}  \\
   \tfrac{w_{k}+E_{T,k}}{2} & r_{k}-f_{s,k}+f_{d,k}  \\
\end{matrix} \right]\geq0.\
\label{eq70}
\end{equation}
Specifically, this form can be extended to accommodate the predefined interval of uncertainties; e.g., when \begin{small}{${\xi }_{k,o}\!\geq\!0$}\end{small}, it is converted to \begin{small}{$\rho_{k}\!\geq\!0$}\end{small} if \begin{small}{$w_{k}\!+\!E_{T,k}\!\geq\!0$}\end{small}, and \begin{small}{$Q_{k}\!+\!\rho_{k}\!\geq\!\sqrt{(w_{k}\!+\!E_{T,k})^{2}\!+\!(Q_{k}\!-\!\rho_{k})^2}$}\end{small} if \begin{small}{$w_{k}\!+\!E_{T,k}\!\leq\!0$}\end{small}, where \begin{small}{$\rho_{k}\!=\!r_{k}\!-\!f_{s,k}\!+\!f_{d,k}$}\end{small}.

Finally, the original problem \eqref{eq59}-\eqref{eq62} is transformed into the following form
\begin{eqnarray}
&&\underset{r_{k},Q_{k},w_{k},y_{k}}{\mathop{\min}}\,r_{k}+y_{k},\
\label{eq71}\\
&&\quad\,{\rm s.t.}\;\left[ \begin{matrix}
   Q_{k} & \tfrac{w_{k}+E_{T,k}}{2}  \\
   \tfrac{w_{k}+E_{T,k}}{2} & r_{k}-f_{s,k}+f_{d,k}  \\
\end{matrix} \right]\geq0,\
\label{eq72}\\
&&\quad\qquad y_{k}\geq{{\gamma}_{k,o2}}{{\Sigma }_{k,o}} Q_{k}+u_{k,o}^{2}Q_{k}+u_{k,o}w_{k}+\sqrt{{\gamma}_{k,o1}}\left|{{\Sigma }_{k,o}^{1/2}}(w_{k}+2Q_{k}u_{k,o})\right|,\
\label{eq73}\\
&&\quad\qquad Q_{k}\geq0.\
\label{eq74}
\end{eqnarray}
In this way, the distributionally robust two-stage model is converted to a tractable form which can be further optimized by the off-the-shelf commercial software.
\subsection{Transformation of Chance Constraints}
In this work, to ensure the availability of power supply with ambiguous PDFs, chance constraints \eqref{eq21} under worst-case uncertain probability distributions should be enforced. That is to say, the chance constraint \eqref{eq21} is recast as the following variant form
\begin{equation}
\begin{aligned}
&{\mathop{\inf}}_{{{\mathbb{P}}_{k,c}}\in {{\mathcal{P}}_{k,c}}}\,\mathbb{P}_{k,c}\big\{-{{\eta }_{T}}{{u}_{f,k}}+D_{e,k}^{u}+D_{e,k}^{s}+{{E}_{T,k}}-{{\eta }_{T}}\tfrac{{{m}_{e,k}}}{{{p}_{ie,k}}}-\tfrac{{{\eta }_{1}}{{\eta }_{G}}}{{{p}_{ig,k}}}{{m}_{g,k}}+\\
&\qquad\qquad\qquad\quad\;{{\eta }_{1}}{{H}_{s,k}}+{{E}_{s,k}}+{{\eta }_{1}}D_{h,k}^{u}+{{\eta }_{1}}D_{h,k}^{s}\le 0\big\}\ge 1-\varepsilon.\
\end{aligned}
\label{eq75}
\end{equation}

However, it is difficult to solve this distributionally robust chance constraint (DRCC) by current optimization tools
in a direct way
due to its' probability type. In this instance, constraint \eqref{eq75} needs to be transformed to a tractable model. The following theorem provides a construction of deterministic counterpart for chance constraints, and the detailed proof can be referred to \cite{zhang2018ambiguous}.
\begin{theorem}\label{th1}
Consider a DRCC ${\mathop{\inf}}_{{{\mathbb{P}}_{k,c}}\in {{\mathcal{P}}_{k,c}}}\,\mathbb{P}_{k,c}\{\bm{\varphi}({\bm{\xi }}_{k,c})^{T}{{\bm{\xi }}_{k,c}}\!\leq\! {{\varphi }^{0}}({\bm{\xi }}_{k,c})\}\geq1\!-\!\varepsilon$,
in which the ambiguity set ${{\mathcal{P}}_{k,c}}$ is defined as \eqref{eq23};  $\bm{\varphi}({\bm{\xi }}_{k,c})$ is the corresponding coefficient vector of ${\bm{\xi }}_{k,c}$; ${{\varphi }^{0}}({\bm{\xi }}_{k,c})\in \mathbb{R}$ represents the limit; and $\varepsilon$ is the risk level of chance constraints.
Then if ${{\gamma }_{k,c1}}/{{\gamma }_{k,c2}}\leq\varepsilon$ this chance constraint is equivalent to
${{\bm{u}}_{k,c}}^{T}\bm{\varphi}({\bm{\xi }}_{k,c})+(\sqrt{{{\gamma }_{k,c1}}}+\sqrt{(1\!-\!\varepsilon)({{\gamma }_{k,c2}}\!-\!{{\gamma }_{k,c1}})/\varepsilon})\sqrt{\bm{\varphi}({\bm{\xi }}_{k,c})^{T}{\bm{\Sigma}}_{k,c}\bm{\varphi}({\bm{\xi }}_{k,c})}\!\leq\!{{\varphi }^{0}}({\bm{\xi }}_{k,c})$; otherwise it equals ${{\bm{u}}_{k,c}}^{T}\bm{\varphi}({\bm{\xi }}_{k,c})\!+\!\sqrt{{{\gamma }_{k,c2}}/\varepsilon}\cdot\sqrt{\bm{\varphi}({\bm{\xi }}_{k,c})^{T}{\bm{\Sigma}}_{k,c}\bm{\varphi}({\bm{\xi }}_{k,c})}\leq{{\varphi }^{0}}({\bm{\xi }}_{k,c})$.
\end{theorem}

According to Theorem \ref{th1}, the DRCC is rewritten as a tractable second-order cone program model
\begin{equation}
\begin{aligned}
&D_{e,k}^{s}+{{E}_{T,k}}-{{\eta }_{T}}\tfrac{{{m}_{e,k}}}{{{p}_{ie,k}}}-\tfrac{{{\eta }_{1}}{{\eta }_{G}}}{{{p}_{ig,k}}}{{m}_{g,k}}+{{\eta }_{1}}{{H}_{s,k}}+{{E}_{s,k}}+{{\eta }_{1}}D_{h,k}^{u}+{{\eta }_{1}}D_{h,k}^{s}+\\
&\qquad\qquad{{\bm{u}}_{k,c}}^{T}[-{{\eta }_{T}}\ 1]^{T}+l_{k}\sqrt{[-{{\eta }_{T}}\ 1]{\bm{\Sigma}}_{k,c}[-{{\eta }_{T}}\ 1]^{T}}\leq0,\
\end{aligned}
\label{eq76}
\end{equation}
where ${{l}_{k}}\!=\!\sqrt{{{\gamma}_{k,c1}}}\!+\!\sqrt{(1\!-\!\varepsilon)({{\gamma}_{k,c2}}\!-\!{{\gamma}_{k,c1}})/\varepsilon}$ if  ${{\gamma }_{k,c1}}/{{\gamma }_{k,c2}}\!\leq\!\varepsilon$; and $l_{k}\!=\!\sqrt{{{\gamma}_{k,c2}}/\varepsilon}$ if ${{\gamma }_{k,c1}}/{{\gamma }_{k,c2}}\!>\!\varepsilon$.
These results have validity with respect to a set of distributions belonging to the ambiguity set ${\mathcal{P}}_{k,c}$, where all PDFs only share the same moment information of uncertain parameters. In addition, the values of the first and second moments are also ambiguous. In this case, we do not have to access to the perfect knowledge of PDFs when dealing with the stochastic program.

\begin{figure}[htbp]
\centering
\advance\leftskip-0.37cm
\includegraphics[height=2.39in,width=6.465in]{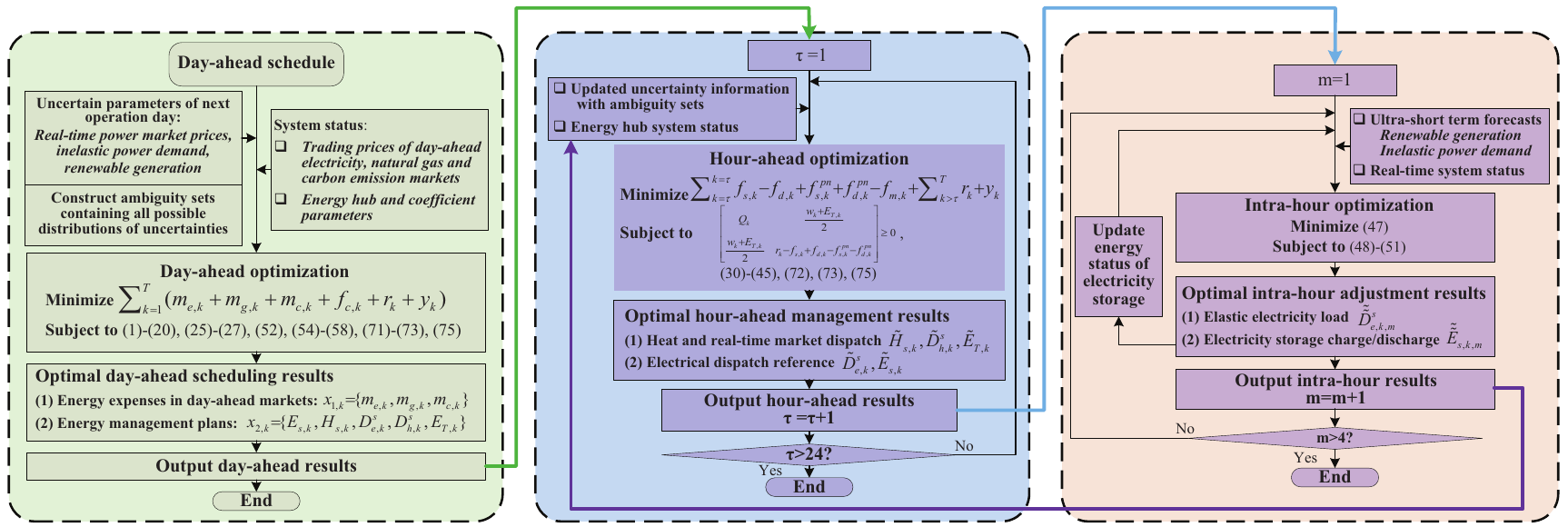}
\caption{Operation diagram of coordinated energy management framework.}
\label{fig4}
\vspace{-0.22cm}
\end{figure}

Therefore through these processes, the day-ahead problem is finally converted to a computationally tractable form which can be efficiently solved by commercial solvers such as CPLEX. The transform method of hour-ahead problem \eqref{eq46} is similar to the above processes, and is omitted here
for the sake of conciseness.
In fact, the proposed coordinated energy management optimization of EH is a multi-phase commit process including the day-ahead level (running every 24 hours), hour-ahead level (running every one hour) and intra-hour level (running every 15 minutes) which is done consequently.
Above all, an operation diagram is illustrated in Fig. \ref{fig4}
to better understand our approach.
\section{Case Study}\label{sec4}
\subsection{Simulation Setup}
\begin{figure}[tbp]
\advance\leftskip-0.5cm
\subfloat[\vspace{-0.18cm}]{
\label{fig5:mini:subfig:a} 
\begin{minipage}[t]{0.33\textwidth}
\centering
\includegraphics[height=1.25in,width=2.19in]{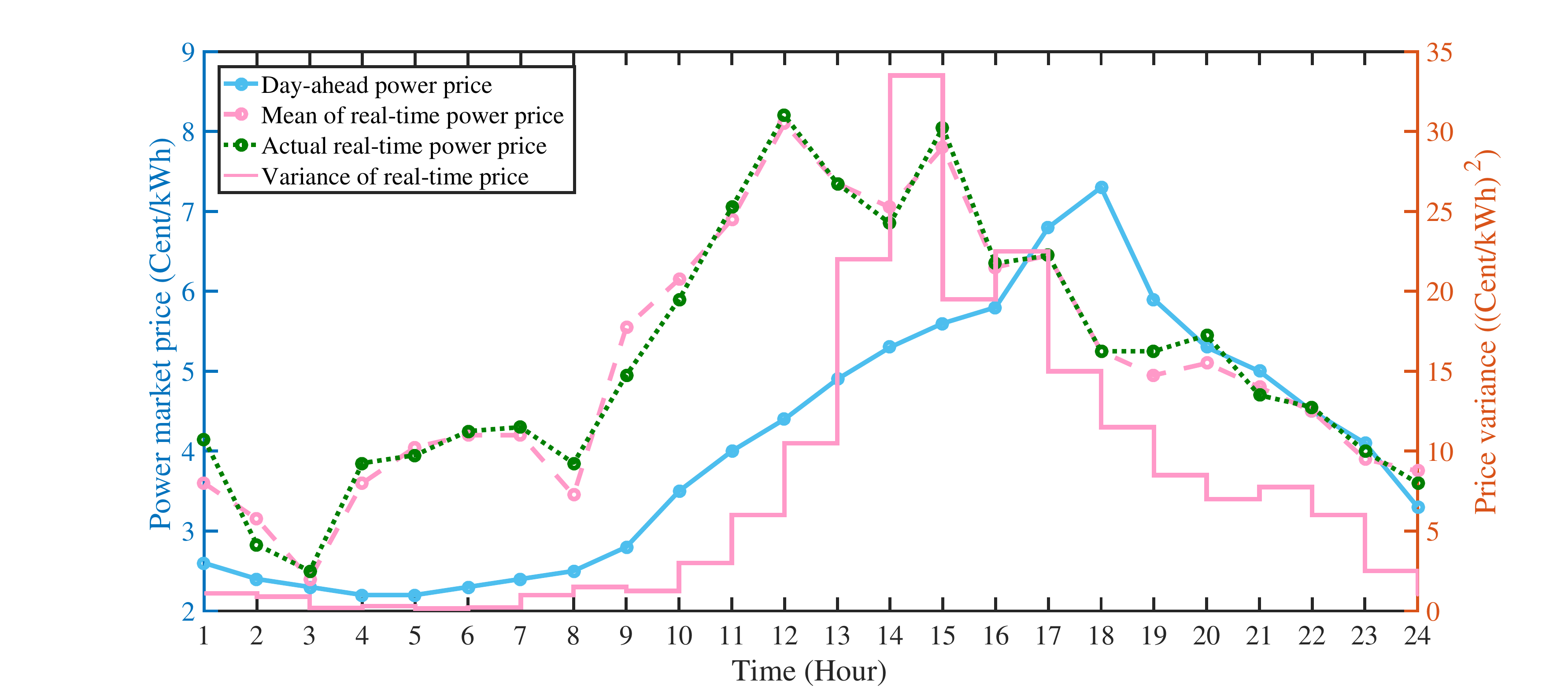}
\end{minipage}}%
\hspace{0cm}
\subfloat[\vspace{-0.18cm}]{
\label{fig5:mini:subfig:b} 
\begin{minipage}[t]{0.33\textwidth}
\centering
\includegraphics[height=1.25in,width=2.2in]{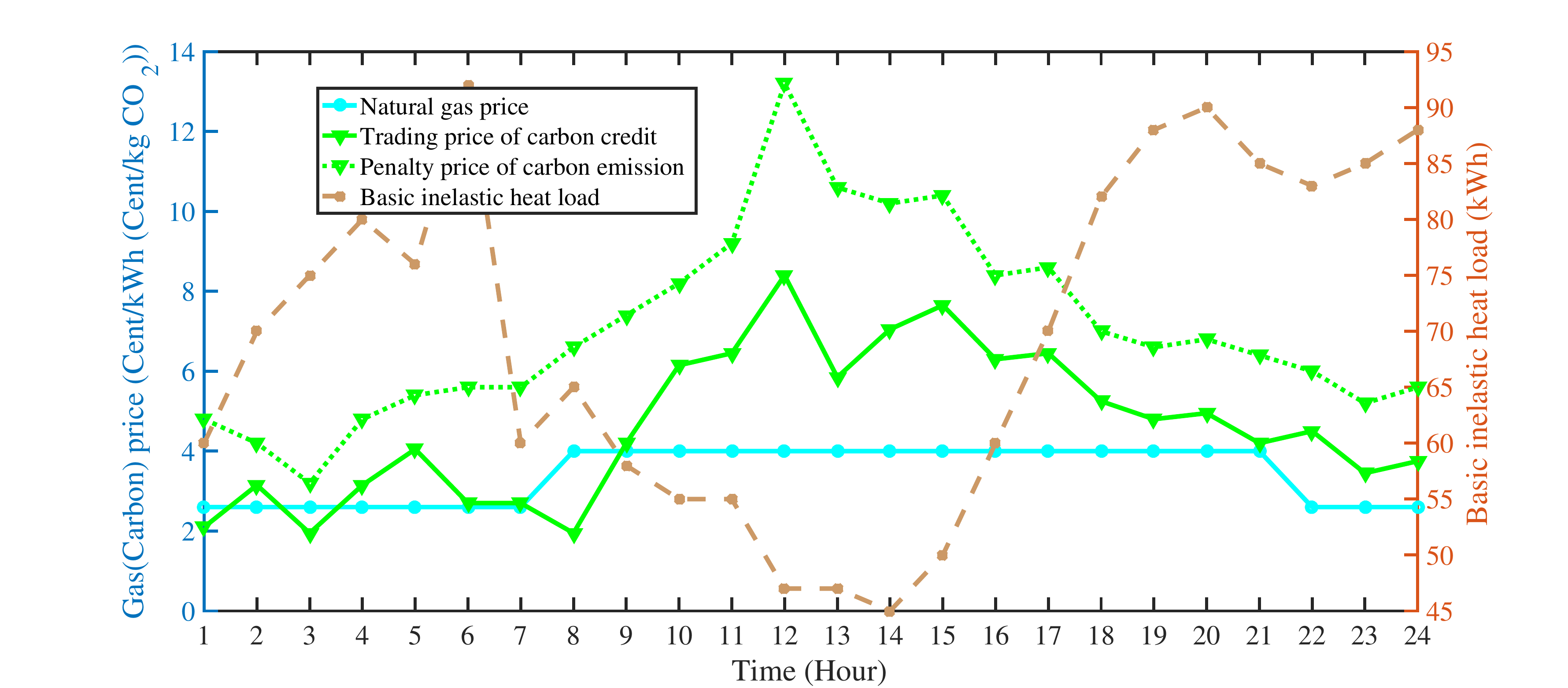}
\end{minipage}}%
\hspace{0cm}
\subfloat[\vspace{-0.18cm}]{
\label{fig5:mini:subfig:c} 
\begin{minipage}[t]{0.33\textwidth}
\centering
\includegraphics[height=1.25in,width=2.2in]{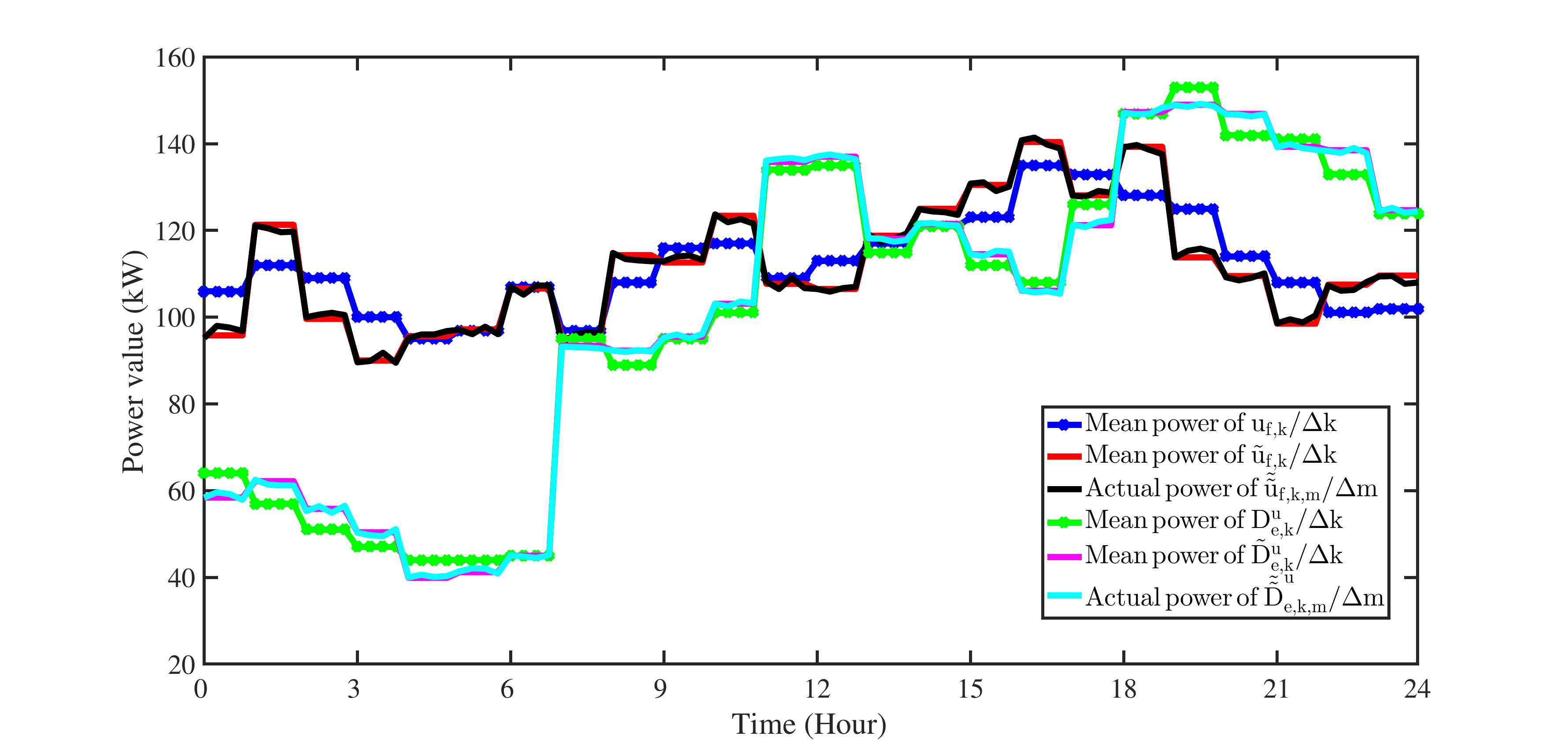}
\end{minipage}}%
\caption{Profiles of market prices, renewables and basic demands.}
\label{fig5:mini:subfig}
\vspace{-0.1cm}
\end{figure}

The goal of this section is to evaluate and discuss the performance of proposed triple-level energy management scheme.
All simulations are implemented on a laptop with 2.50 GHz, Intel Core i5-7200 CPU using MATLAB with YALMIP toolbox and CPLEX 12.6.3.
Primary equipment and system parameters are summarized in Table \ref{Tab1}.
Profiles of 24-h energy data containing day-ahead power prices, predicted mean and actual real-time power prices, price variances (displayed in Fig.\ref{fig5:mini:subfig}~\subref{fig5:mini:subfig:a}), gas prices, trading prices of carbon credit, penalty prices of extra carbon emissions and inelastic heat load (drawn in Fig.\ref{fig5:mini:subfig}~\subref{fig5:mini:subfig:b}) are extracted and tailored from \cite{liveprice,aboli2019joint,javadi2020optimal}.
In Fig.\ref{fig5:mini:subfig}~\subref{fig5:mini:subfig:c} the mean/actual power value of renewable output and basic electricity load over different timescales are illustrated. The risk level $\varepsilon$ of chance constraint is set to 0.05.
\begin{table}[ht]
\centering
\fontsize{5}{2.5}\selectfont
\caption{Parameters of EH and system operations.\vspace{-0.15cm}}
\setlength{\tabcolsep}{0.81mm}{%

\begin{tabular}{@{}cccccccccccc@{}}

\toprule

Parameter&Value& Parameter & Value & Parameter & Value & Parameter & Value & Parameter & Value & Parameter & Value\\[1pt] \midrule
$\eta_{T}$ &  0.98 &  $\eta_{M}^{e}$ &  0.35 &  $\eta_{M}^{h}$ &  0.4 &  $\eta_{G}$ &  0.9 & $\eta_{E}^{ch(dis)}$ &  0.97 &  $\eta_{H}^{ch(dis)}$ &  0.87\\[1pt] \midrule
$E_{in}^{\max}$(kWh)& 130 & $G_{in}^{\max}$(kWh) & 350 & $E_{T}^{\max}$(kWh) &  100 &  $G_{MT}^{\max}$(kWh) &  300 &  $G_{GF}^{\max}$(kWh) &  250 &  $G(E)_{in}^{\min}$(kWh) &  0\\[1pt] \midrule
$E_{s,k}^{ch(dis),\max}$(kW)& 20 & $H_{s,k}^{ch(dis),\max}$(kW) & 50 & $S_{E,k}^{\min}$(kWh) &  20 &  $S_{E,k}^{\max}$(kWh) &  160 &  $S_{H,k}^{\min}$(kWh) &  50 &  $S_{H,k}^{\max}$(kWh) &  500\\[1pt] \midrule
\tabincell{c}{$C_{o2e}$\\(kg$\rm{CO_{2}}$/kWh)}& 0.6 & \tabincell{c}{$C_{o2g}$\\(kg$\rm{CO_{2}}$/kWh)} & 0.5 & \tabincell{c}{$\eta_{k}P_{lf,k}$\\(kg$\rm{CO_{2}}$)} &  110 &  \tabincell{c}{$\alpha_{e(h),k}$\\(\textcent/$\rm{kWh^{2}}$)} &  -0.08 &  \tabincell{c}{$\beta_{e,k}$\\(\textcent/$\rm{kWh}$)} &  8 &  \tabincell{c}{$\beta_{e,k}$\\(\textcent/$\rm{kWh}$)} &  7\\[1pt]
\bottomrule

\end{tabular}%

}
\label{Tab1}
\vspace{-0.39cm}
\end{table}
\subsection{Results and Analysis}
\subsubsection{Optimal results of triple-layered energy management}
The energy management results of EH based on the triple-layered timescale structure and proposed optimization approach are shown in this subsection with the above given test parameters.
\begin{figure}[htbp]
\centering
\includegraphics[height=1.7in,width=5in]{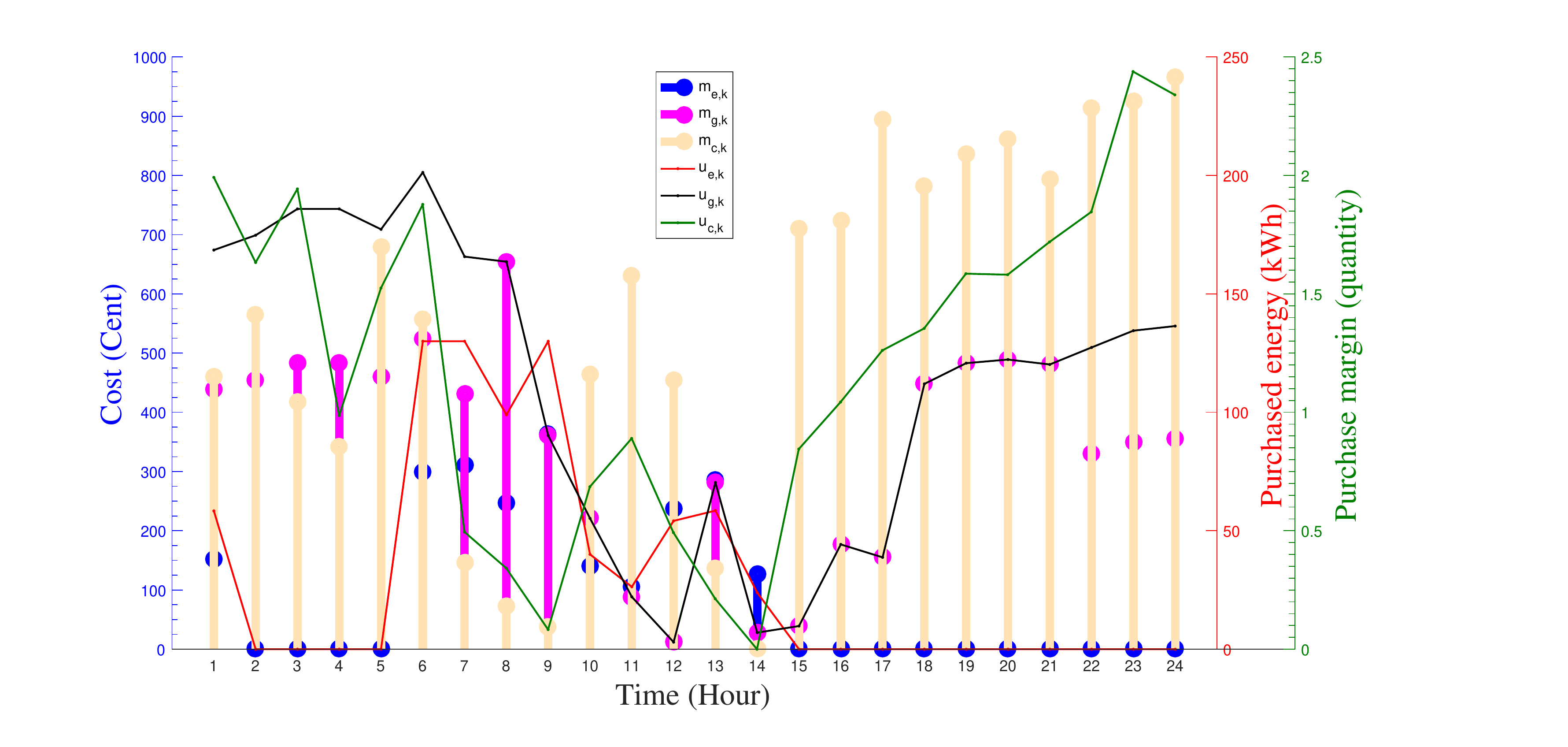}
\caption{Optimal results of energy purchase.}
\label{fig6}
\vspace{-0.36cm}
\end{figure}

\begin{figure}[tbp]
\advance\leftskip-0.28cm
\subfloat[\vspace{-0.18cm}]{
\label{fig7:mini:subfig:a} 
\begin{minipage}[t]{0.5\textwidth}
\centering
\includegraphics[width=3in]{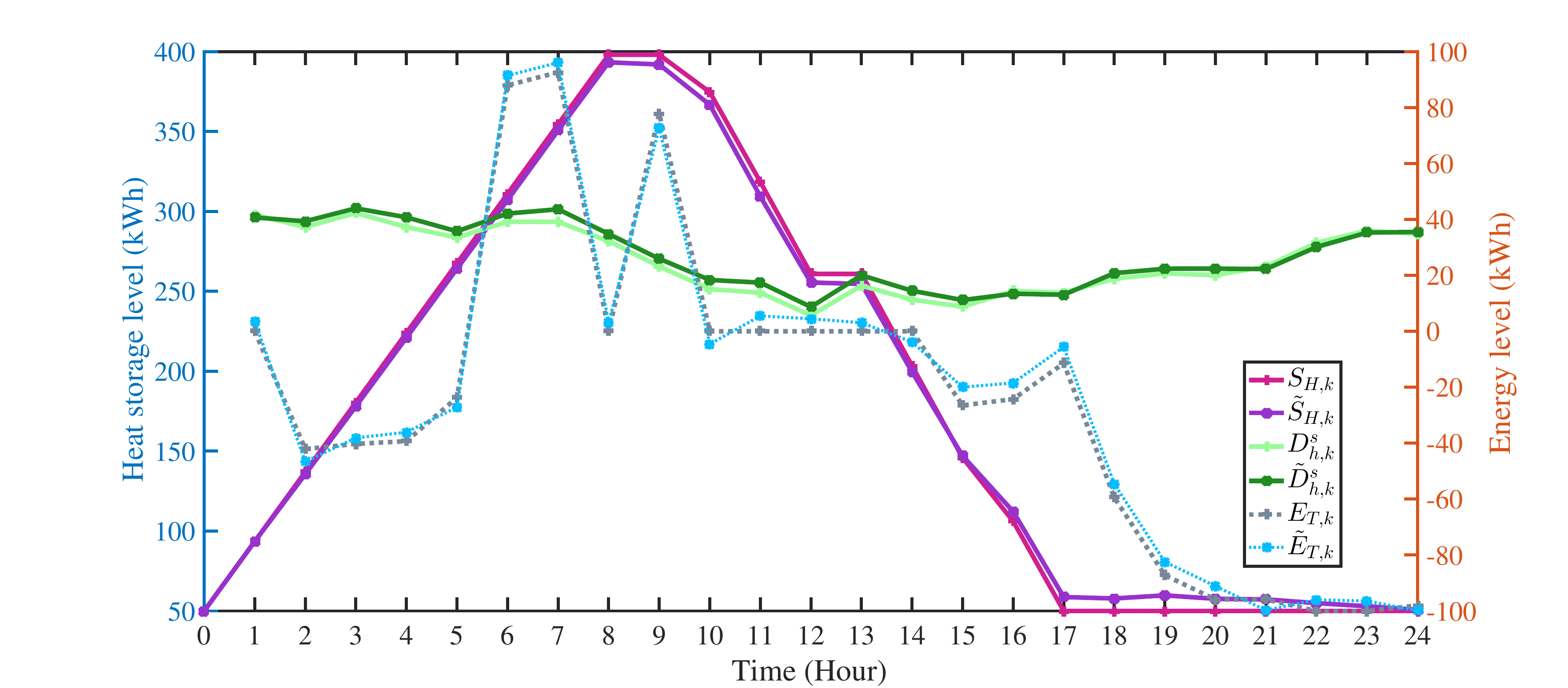}
\end{minipage}}%
\hspace{0cm}
\subfloat[\vspace{-0.18cm}]{
\label{fig7:mini:subfig:b} 
\begin{minipage}[t]{0.5\textwidth}
\centering
\includegraphics[width=3in]{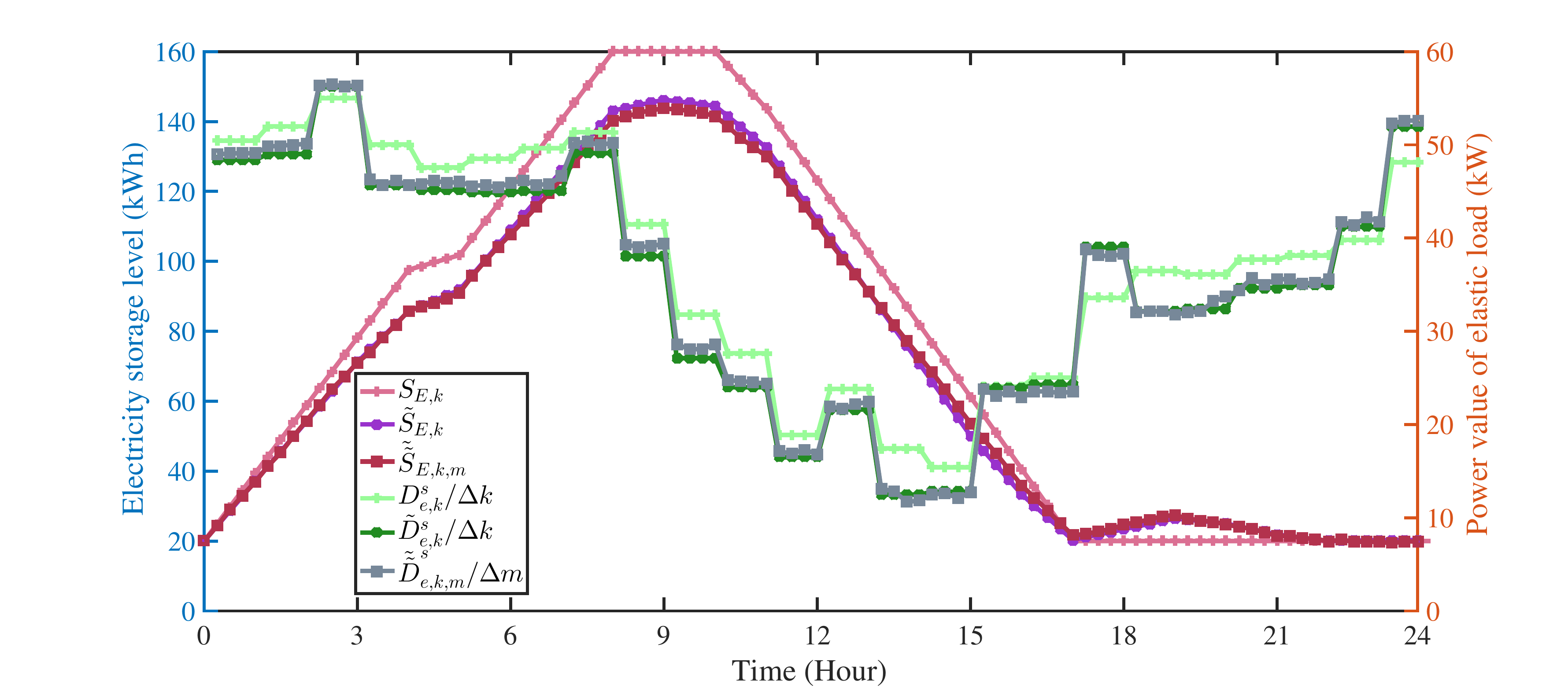}
\end{minipage}}%
\caption{Optimal scheduling results of storages and demands.}
\label{fig7:mini:subfig}
\vspace{-0.1cm}
\end{figure}

Fig.\ref{fig6} demonstrates the corresponding bidding cost with stem chart, purchased energy and carbon purchase margin from the power company, gas company and emission trading center in day-ahead energy markets with line charts.
Fig.\ref{fig7:mini:subfig}\subref{fig7:mini:subfig:a} and Fig.\ref{fig7:mini:subfig}\subref{fig7:mini:subfig:b} present the optimized energy scheduling results in the day-ahead and intra-day rolling horizon with hour-ahead and intra-hour timescales, respectively.
From Fig.\ref{fig6}, it is seen that the purchased electricity is small in several slots. There are mainly two reasons for this phenomenon.
On one hand, the price of natural gas is usually cheaper than that of power market, which implies more economical.
On the other hand, recall that power purchased from the day-ahead wholesale market is generated by coal-fired units the carbon intensity of which are higher than natural gas.
Besides, renewables are carbon-free sources, and as mentioned in Section \ref{sec221}, the energy transaction in real-time power market from other producers is assumed to be produced by renewable generators.
Therefore it motivates EH most likely to select inexpensive or low-carbon energy resources.
Without changing day-ahead energy purchase strategies in Fig.\ref{fig6},
EH, through the cooperation between hour-ahead and intra-hour adjustments as indicated in Fig.\ref{fig7:mini:subfig}, can gradually re-regulate heat storages, elastic heat loads, real-time power market transactions in hour-ahead phases, and electricity storages, elastic power loads during intra-hour phases to smooth out uncertainty fluctuations. Meanwhile, the results can track the upper-level scheduling as far as possible ensuring the effectiveness of economic dispatch.
In addition, the final energy state of heat storage is observed to be equivalent to the original level (\begin{small}{$50$kWh}\end{small}). The final electricity storage status during the intra-hour stage also returns to its pre-set initial value (\begin{small}{$20$kWh}\end{small}) which indicates a cyclic energy regulation is ensured for the next scheduling period.
It can be observed that the intra-day hour-ahead scheduling has played a `buffer conditioning' role in the proposed triple-layered structure correlating the day-ahead scheduling with intra-hour dispatches.
Furthermore, based on the proposed day-ahead schedule optimization, the intra-day two-level adjustments can efficiently accommodate the variations in real-time market prices and fluctuations in renewables and power load step by step.
The average solution times of day-ahead, hour-ahead and intra-hour level optimization are respectively $69.4$s, $15.3$s, $5.1$s, which are enough to match the operation performance requirement of day-ahead and intra-day scheduling compared with their individual timescale intervals.
\subsubsection{Impacts of uncertain parameters}\label{sec722}
\begin{figure}[htbp]
\advance\leftskip-0.5cm
\subfloat[\vspace{-0.18cm}]{
\label{fig8:mini:subfig:a} 
\begin{minipage}[t]{0.33\textwidth}
\centering
\includegraphics[height=1.3in,width=2.2in]{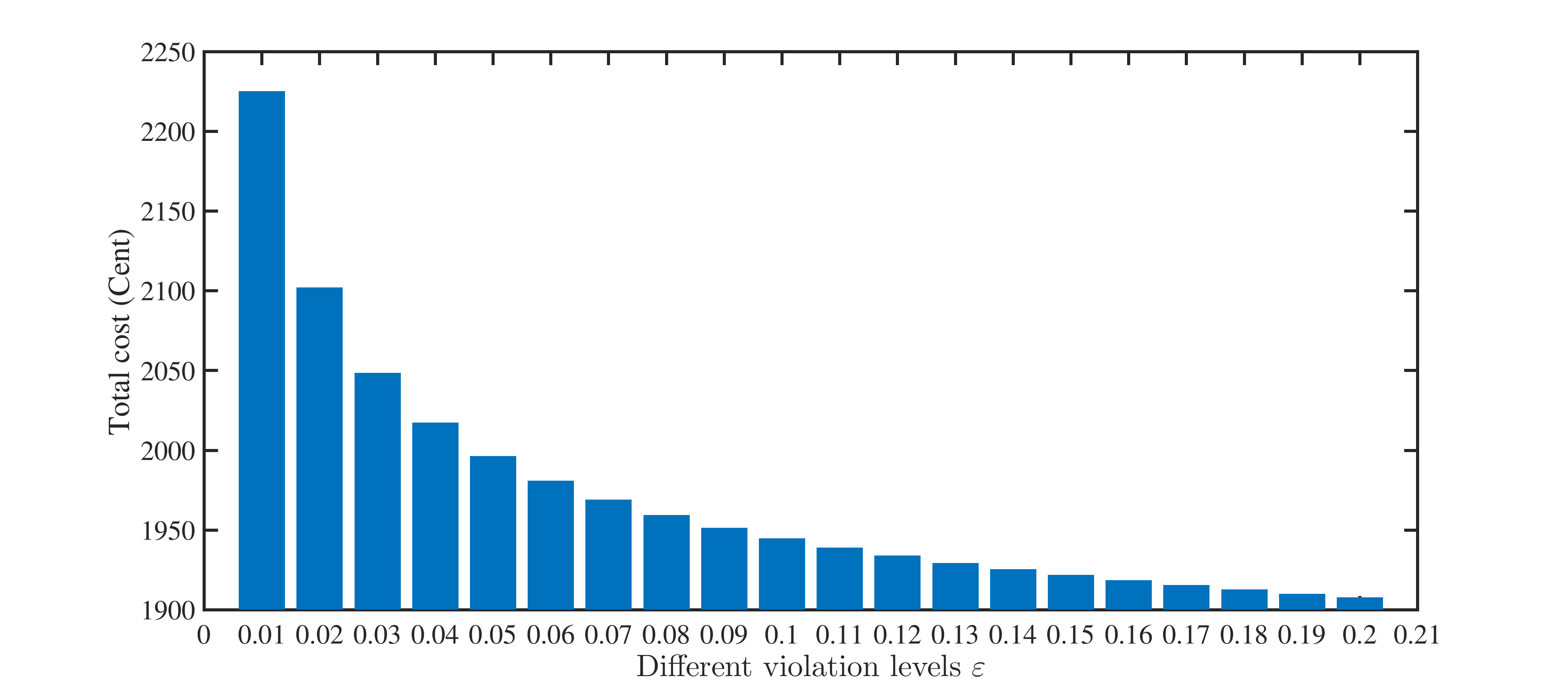}
\end{minipage}}%
\hspace{-0.15cm}
\subfloat[\vspace{-0.18cm}]{
\label{fig8:mini:subfig:b} 
\begin{minipage}[t]{0.33\textwidth}
\centering
\includegraphics[height=1.3in,width=2.2in]{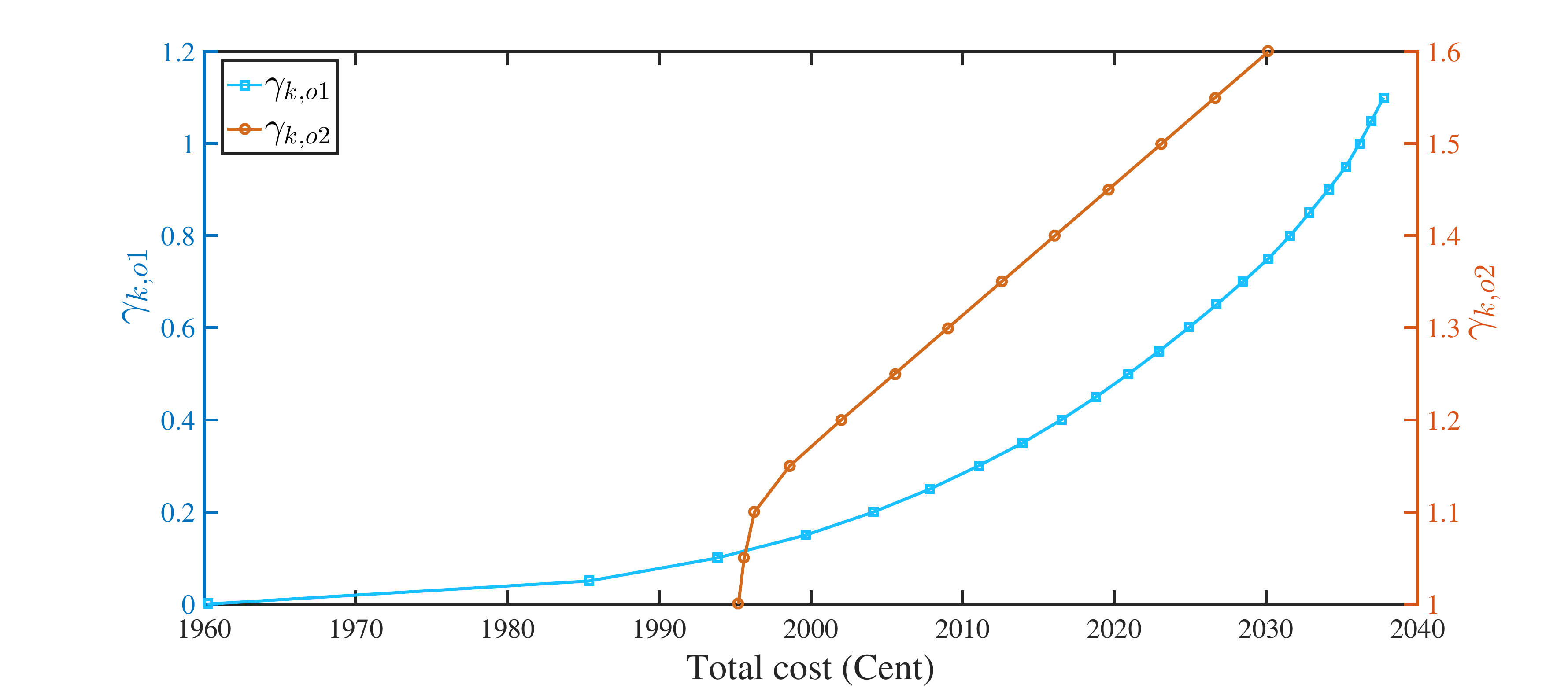}
\end{minipage}}%
\hspace{-0cm}
\subfloat[\vspace{-0.18cm}]{
\label{fig8:mini:subfig:c} 
\begin{minipage}[t]{0.33\textwidth}
\centering
\includegraphics[height=1.3in,width=2.2in]{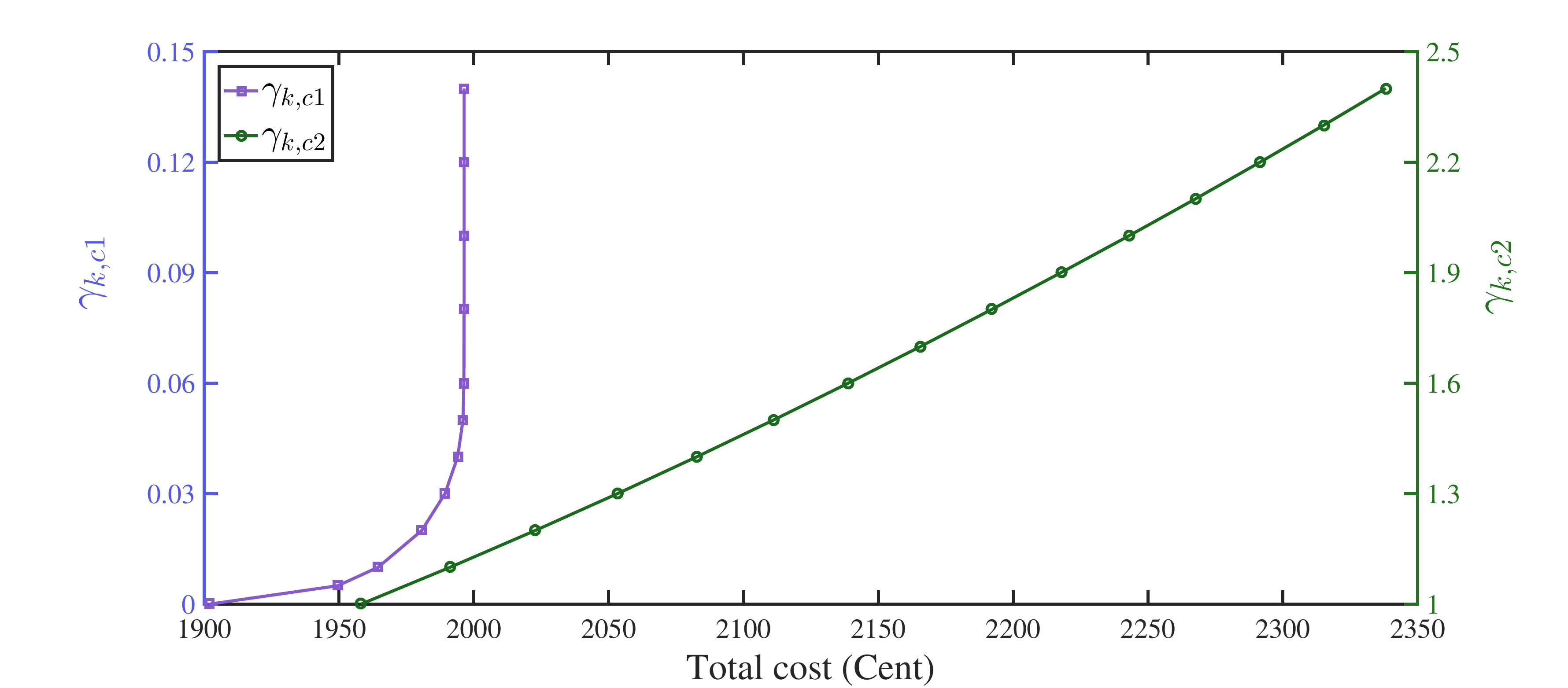}
\end{minipage}}%
\caption{Total costs with changing parameters.}
\label{fig8:mini:subfig}
\vspace{-0.1cm}
\end{figure}

Sensitivity analyses are further carried out to test the energy management performance of the proposed model with different parameters. The results are presented in Fig.\ref{fig8:mini:subfig}. First, the impacts of confidence levels $1-\varepsilon$ on total costs are investigated aside from the usually adopted one $95\%$ with different violation probability values $\varepsilon$ from $0.01$ to $0.2$ with a step of $0.01$. It is seen from Fig.\ref{fig8:mini:subfig}~\subref{fig8:mini:subfig:a} that the total cost increases as $\varepsilon$ becomes lower (i.e., confidence levels get larger) to pay more heed to system reliability. In addition, a rapid increase in total cost is also observed around $\varepsilon\in[0.01\;0.04]$.
It implies that with a further decrease in risk level to a certain extent, marginal costs may rise sharply. This phenomenon illustrates that a compromise between the rigorous reliability and economic benefit can be prescribed and regulated by the confidence level which is favorable for practical applications.

Following that, impacts of moment uncertain size concerning ambiguity info about means and variances in ${{\xi }_{k,o}}$ $({{\gamma}_{k,o1}},{{\gamma}_{k,o2}})$ and ${{\bm{\xi }}_{k,c}}$ $({{\gamma}_{k,c1}},{{\gamma}_{k,c2}})$ on total costs are studied by varying one parameter at a time and holding others fixed. In Fig.\ref{fig8:mini:subfig}~\subref{fig8:mini:subfig:b}, given ${{\gamma}_{k,c1}}\!=\!0.12,{{\gamma}_{k,c2}}\!=\!1.12,\varepsilon\!=\!0.05$, trends of total costs with changing ${{\gamma}_{k,o1}}$, ${{\gamma}_{k,o2}}$ are plotted as the blue line and brown line respectively.
It is observed that total costs get larger when a higher ${{\gamma}_{k,o1}}$ or ${{\gamma}_{k,o2}}$ is arrived at. In other words, the more ambiguous the moment information of PDFs, the higher total cost is procured to sustain the system operation at a lower risk level.
Besides, increments of costs with varying ${{\gamma}_{k,o1}}$ and ${{\gamma}_{k,o2}}$ have shown
different growth tendencies because they are based on distinct centered moment values.
In this regard, these moment size parameters shall be appropriately approximated and decreased by some data-driven techniques based on the mass data \cite{delage2010distributionally}, which is still left open for further exploration.
Similarly, with fixed ${{\gamma}_{k,o1}}\!=\!0.12,{{\gamma}_{k,o2}}\!=\!1.12,\varepsilon\!=\!0.05$, the varieties of total costs under different ${{\gamma}_{k,c1}}$, ${{\gamma}_{k,c2}}$ are indicated separately as the purple and green lines shown in Fig.\ref{fig8:mini:subfig}~\subref{fig8:mini:subfig:c}.
A higher total cost is also observed with increasing the value of ${{\gamma}_{k,c1}}$ or ${{\gamma}_{k,c2}}$. Particularly, when ${{\gamma}_{k,c1}}$ reaches around 0.06, the cost remains almost unchanged. This is attributed to the fact ${{\gamma }_{k,c1}}/{{\gamma }_{k,c2}}\!>\!\varepsilon$ in \eqref{eq76} held.
\subsubsection{Impacts of carbon trading}
\begin{figure}[htbp]
\centering
\includegraphics[width=6.2in]{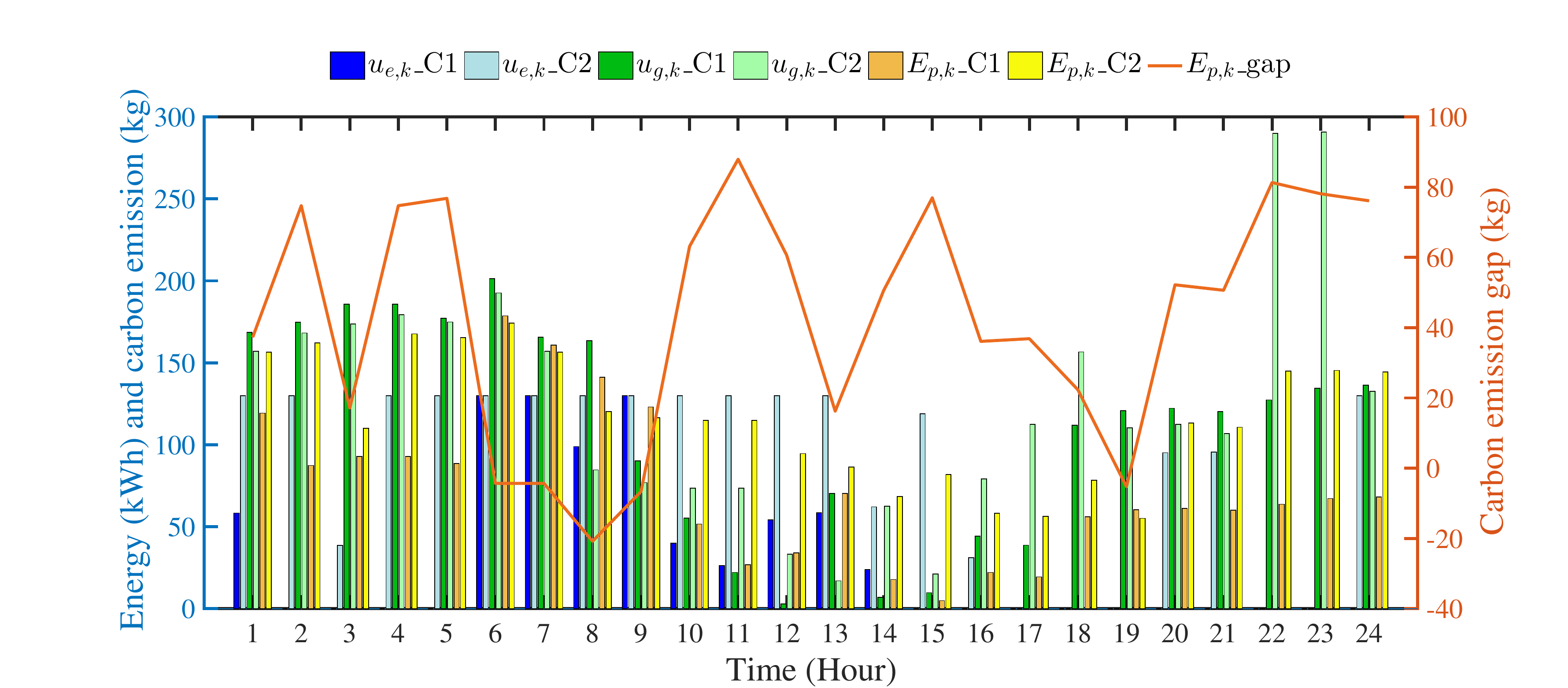}
\caption{Performance of carbon trading.}
\label{fig9}
\vspace{-0.36cm}
\end{figure}

Recently carbon trading has received growing attention in energy consumption. Fig.\ref{fig9} has presented the differences in energy purchase and carbon emissions of an EH based on the proposed approach with the same total budget inputs by employing the following two market cases: C1 is the employed model with carbon trading market; as opposed to C1, C2 omits carbon trading while other market components are unchanged.
It is observed that the purchased electricity $u_{e,k}$ from C1 is always lower than that from C2 in each slot owing to a higher carbon intensity of coal-fired power. And C1 produces a $64\%$ decrease in the total purchased electricity compared to the result of C2. As for the purchased natural gas $u_{g,k}$, the result of C1 ($u_{g,k}$\_C1) is usually higher than that of C2 ($u_{g,k}$\_C2) in time slots 1-9, 13, 19-21 and 24. Except for the above periods, $u_{g,k}$\_C2 procures a bigger volume because that it does not pay for carbon emissions and consequently has excess money to purchase gas.
There is a reduction of $13\%$ for the purchased natural gas $u_{g,k}$\_C1 as compared to the volume of $u_{g,k}$\_C2.
In addition, lower emissions are almost observed in C1
where the summation of $E_{p,k}\_\rm{C1}$ is reduced by $37\%$ relative to the one of C2.
It implies that participating carbon trading market has a significant influence on energy purchase strategies (i.e., tend to support low-carbon energy resources) accompanied with promoting the carbon emission reduction subsequently.
\subsubsection{Impacts of multi-timescale optimization}
%
%
%
%
In this subsection, several models are illustrated to analyze the performance of proposed intra-day two-timescale optimization. Table~\ref{Tab2} has reported the corresponding total cost of an EH in  hour-ahead level and intra-hour level under four different intra-day dispatches denoted as \begin{small}{M1-M4}\end{small}. \begin{small}{M1}\end{small}, as the basic case, is the proposed two-timescale coordination model with a receding horizon where multi-level re-regulation is performed based on the results of previous phase and updated information of system state. As for \begin{small}{M2}\end{small}, there is only an intra-hour level for the intra-day realtime dispatch where the scheduling is directly based on the day-ahead DRO results without any hour-ahead buffering and revised regulations.
Consequently, the total cost of intra-hour level has increased by \begin{small}{$53.8$\textcent}\end{small} compared with \begin{small}{M1}\end{small} as shown in Table~\ref{Tab2}.
In \begin{small}{M3}\end{small}, the constraints of final electricity storage state are ignored and the intra-hour storage statuses are not inserted into hour-ahead dispatches.
To that point, the initial electricity storage state is not ensured for the following receding energy management optimization, and over multi-period the state of storage may arrive at the lowest energy level with a subsequent considerable rise in total operation costs.
In addition, the storage is charged and discharged freely owing to the unlimited final storage state.
However, the storage state may not meet the initial required value for the next day-ahead dispatch to the detriment of healthy cyclic storage regulation.
\begin{small}{M4}\end{small} omits feedbacks regarding the intra-hour optimized elastic demand on the scheduling in hour-ahead phases, and thereby with the tapering regulation ability and bigger operation cost. In consequence of missing real demand updates in each period, the cost gap between hour-ahead and intra-hour levels is slightly increased from \begin{small}{$12.1$\textcent}\end{small} in \begin{small}{M1}\end{small} to \begin{small}{$15.4$\textcent}\end{small} in \begin{small}{M4}\end{small}.
The results suggest that the proposed intra-day optimization model incorporating multi-timescale with receding horizon may have the ability to improve economic operation with circular scheduling.
\begin{table}[t]
\centering
\fontsize{6.5}{4}\selectfont
\caption{Performance of multi-timescale framework (given unit: cent).\vspace{-0.15cm}}
\setlength{\tabcolsep}{4.9mm}{%

\begin{tabular}{lllll}

\toprule
Model & M1 & M2 & M3 & M4\\[2pt] \midrule
\tabincell{c}{Total cost in hour-ahead level}&  2058.3 &  $\cdots$ &  2118.7 &  2067.5\\[2pt] \midrule
\tabincell{c}{Total cost in intra-hour level}& 2070.4 & 2124.2 & 2130.6 & 2082.9\\[2pt] \bottomrule

\end{tabular}%

}
\label{Tab2}
\vspace{-0.38cm}
\end{table}
\subsubsection{Comparisons of economic performance}
In this subsection, economical results in the triple-layered framework are further evaluated to investigate the effectiveness of proposed two-stage DRO scheduling approach (TSDRO) with three alternative methods as presented in Table~\ref{Tab3}.
We first offer a brief introduction of these methods: 1) As for traditional two-stage stochastic programming method (TSSP) with a general mathematical model $\{\min_{\bm{x}_{1}}\bm{c}^{\rm T}{\bm{x}_{1}}\!+\!\mathbb{E}_{\bm{\xi}}[\min_{\bm{x}_{2}}\bm{b}^{\rm T}(\bm{\xi})\bm{x}_{2}(\bm{\xi})]: \bm{A}\bm{x}_{1}\!\geq\!\bm{d},\;\bm{B}(\bm{\xi})\bm{x}_{1}\!+\!\bm{E}(\bm{\xi})\bm{x}_{2}(\bm{\xi})\!\geq\!\bm{h}(\bm{\xi})\}$ where $\bm{x}_{1},\bm{x}_{2}$ two-stage decision variables, $\bm{\xi}$ uncertain parameters, $\bm{c},\bm{b},\bm{A},\bm{d},\bm{B},\bm{E},\bm{h}$ the corresponding coefficient matrices in objective and constraints,
it is usually a scenario-based form with predefined exact scenario probabilities employing Monte Carlo simulation.
2) As for two-stage robust optimization (TSRO) with a standard form $\{\min_{\bm{x}_{1}}\bm{c}^{\rm T}{\bm{x}_{1}}\!+\!\max_{\bm{\xi}\in \mathbb{U}}\min_{\bm{x}_{2}}\bm{b}^{\rm T}\bm{x}_{2}: \bm{A}\bm{x}_{1}\!\geq\!\bm{d},\;\bm{E}\bm{x}_{2}\!\geq\!\bm{h}\!-\!\bm{B}\bm{x}_{1}\!-\!\bm{C}\bm{\xi}\}$
where uncertain parameters $\bm{\xi}$ are described by the interval uncertainty sets, it can be efficiently solved by column-and-constraint generation technique and Benders decomposition algorithms.
More information and detailed solution techniques about TSSP and TSRO can be referred to 
\cite{al2021two,zhao2018strategic,7307233,aboli2019joint}.
3) The deterministic optimization method (DOM) does not consider any uncertainties in day-ahead scheduling.
%
%
%
\begin{table}[htbp]
\centering
\fontsize{6.5}{6}\selectfont
\caption{Cost comparisons under different methods (given unit: cent).\vspace{-0.15cm}}
\setlength{\tabcolsep}{2.2mm}{%

\begin{tabular}{ccccccc}

\toprule
Method &\tabincell{c}{DA energy purchase$^a$\\and carbon cost}&\tabincell{c}{Energy storage cost\\and demand utility}&\tabincell{c}{Average RT power$^b$\\market trading cost}&\tabincell{c}{Total cost\\in DA level}&\tabincell{c}{Total cost$^c$\\in HA level}&\tabincell{c}{Total cost$^d$\\in IH level}\\[2pt] \midrule
TSDRO&  5538.8 &  -6867.7 &  3325.3 &  1996.4 &  2058.3 &  2070.4\\[2pt]
TSSP&  5521.3 &  -6880.2 &  3298.4 &  1939.5 &  2043.8 &  2083.6\\[2pt]
TSRO&  5576.1 & -6840.1 &  3339.2 &  2075.2 &  2124.1 &  2138.5\\[2pt]
DOM& 5493.8 & -6898.4 & 3265.2 & 1860.6 &  2107.2 &  2194.9\\[2pt] \bottomrule
\end{tabular}%

}
\label{Tab3}
\tiny{$^a$ DA: day-ahead, $^b$ RT: real-time, $^c$ HA: hour-ahead, $^d$ IH: intra-hour}\\
\vspace{-0.38cm}
\end{table}

Table \ref{Tab3} has reported six types of system costs corresponding to the aforementioned methods. As can be observed from Table \ref{Tab3}, TSRO considering the worst-case scenario but not the distribution probability of uncertainties in the second stage form procures subsequent bigger system operation cost (i.e., energy transaction cost and utilities) and total cost in day-ahead management.
The results are more conservative (robustness) than those in the proposed TSDRO and TSSP. However, the average adjacent cost gaps among day-ahead, hour-ahead and intra-hour levels are much smaller than TSSP and DOM.
In TSSP, expected costs with known distribution probabilities are considered in the second stage form of day-ahead problem and bring about a corresponding lower system cost.
Similar to TSSP, DOM exhibits a least day-ahead cost but does not take account of any uncertainties. Although intra-day re-regulations may amend the day-ahead scheduling, the robustness of deterministic model is too deficient to hedge against uncertainties. Therefore total costs of intra-hour level are largest. As to TSDRO, its day-ahead and hour-ahead costs are higher than those of TSSP but lower than those of TSRO because its two-stage optimization can be deemed as a combination of SP and RO
in some sense.
Moreover, the total intra-hour level operation costs of TSDRO accommodating fluctuations in uncertain parameters can be reduced averagely 3\% compared with other three contrasting cases owing to a more comprehensive consideration of probability distribution uncertainties in the previous day-ahead and hour-ahead scheduling.
It indicates that TSDRO sacrificing some robustness in exchange for a less conservative strategy is still robust enough to handle uncertain scenarios in practice owing to the considering of
uncertain probability distributions.
\subsubsection{Scalability analysis}
\begin{figure}[htbp]
\centering
\includegraphics[height=2.02in,width=4.5in]{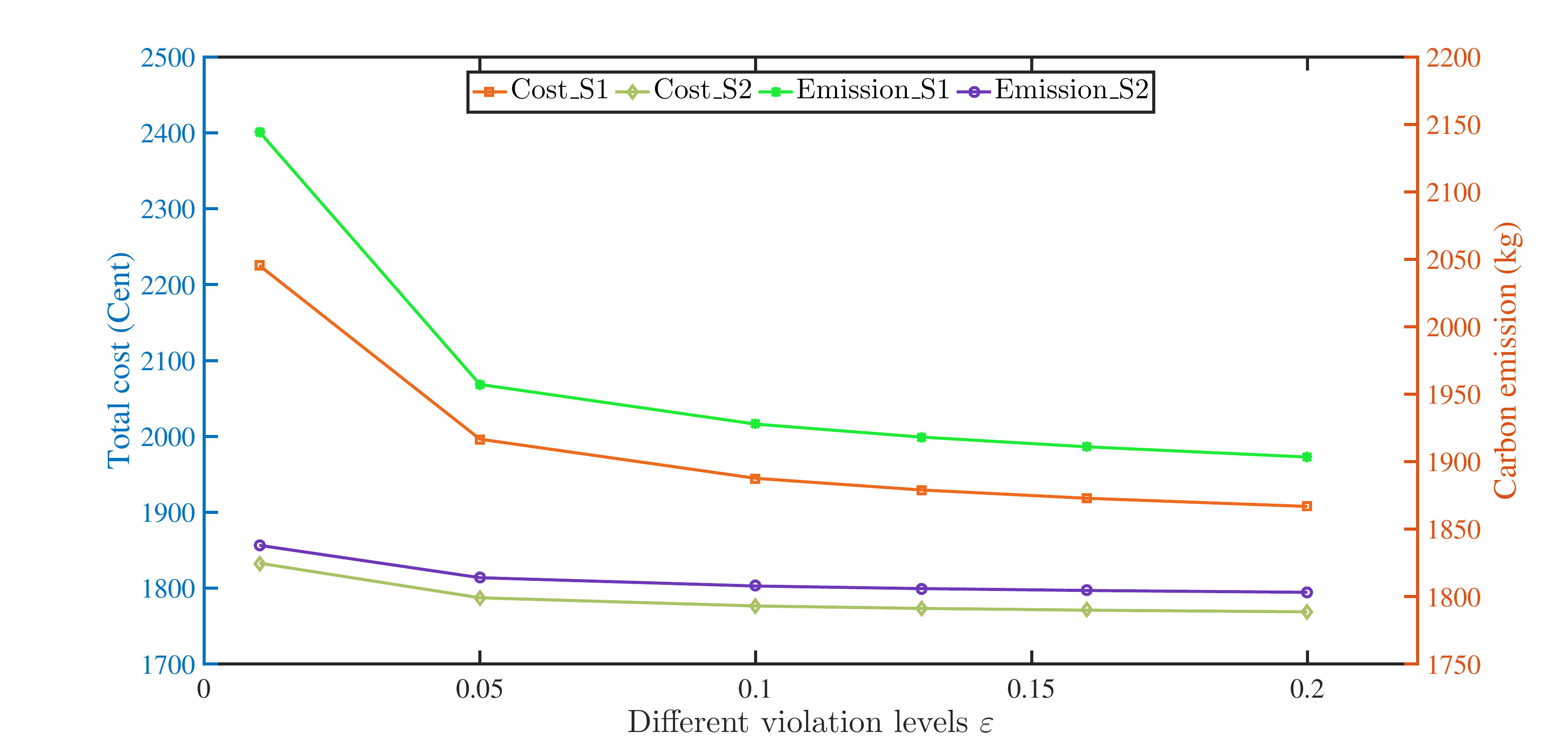}
\caption{Performance results of a larger EH.}
\label{fig10}
\vspace{-0.26cm}
\end{figure}
In this subsection, the proposed method is also implemented on a larger EH system to verify the scalability.
Compared to the benchmark EH system in previous sections denoted as S1, the new EH denoted as S2 applies the quintuple number of conversion devices (i.e., gas furnaces, micro turbines and RESs). The primary simulation setups of S2 are the same as those in S1. In addition, S2 also includes the hydrogen generation plant, hydrogen storages and fuel cell plant as other controllable resources. The model formulations and parameter settings of these controllable resources can be found in our previous work~\cite{9420353}. Fig.\ref{fig10} has shown the comparison results on the total costs and carbon emissions with varying  violation probability values (confidence levels $1-\varepsilon$).
It is observed that the total costs of both systems are increased with the decline of $\varepsilon$ seeking for a more rigorous system reliability. This phenomenon is consistent with that of previous Section~\ref{sec722}, which demonstrates the effectiveness of the potential application on a larger EH.
Similarly, the uptrend of carbon emissions with a smaller $\varepsilon$ can be observed. This is because to maintain the strict system reliability, EH tends to pursue more energy resources and produce emissions.
Besides, compared with the results of S1,  S2 has always shown lower total costs and carbon emissions.
The main reason is that S2 has more renewable supply and gas conversion devices which lead to a shrunken need for the external power supply. Recall that the renewables are assumed to be carbon-free resources, and the carbon intensity in natural gas conversion is usually smaller than that of the external coal-fired electricity generation.
In addition, the hydrogen conversion devices enable S2 to gain more operation feasibility, which may avoid purchasing certain electricity from markets at the peak price periods.
Moreover, similar operation results demonstrated in S1 can also be attained for S2, the details of which are omitted here.
Furthermore, the average computation time of S2 is about 79.8s which is acceptable for practical application.
Above all, the performance results validate that the proposed method is scalable and has potential applications for extensible case studies.
\section{Conclusion}\label{sec5}
The coordinated energy management problem of an EH in hybrid markets is investigated considering uncertain renewables, power demand and real-time market prices.
We model this problem as a multi-phase framework to jointly optimize the triple-level scheduling results, which is done by developing CC-based two-stage DRO models and the multi-timescale scheme with a receding horizon. Energy imbalances resulted from uncertainties are restrained by chance constraints with worst-case distributions, and provided by intra-day scheduling eventually.
Relevant reformulation processes are given to obtain a computationally tractable form.
Numerical studies demonstrate that the proposed model can provide a less conservative referral plan for EH.
By the comparison with other models, the effectiveness and flexibility of the proposed model in terms of the tradeoff between optimality and robustness, carbon emissions are verified.
In the future, the developed framework will extend to consider dependent uncertainties where energy market prices are motivated by dispatch strategies, available resources, and to analyze their impacts on social welfare.

In the paper, we utilize a linear multi-energy coupling model to express the heat-power conversion relationship.  It is articulate and convenient for the formulation. Due to the concern of suboptimality/infeasibility, a more realistic scenario is that the heat-power relationship is not linear with variable conversion ratios. This is a significant research direction and deserves further investigation.
Besides, in view of modern smart energy and telecommunication infrastructures, the energy scheduling is vulnerable to cyberattacks such as injecting false real-time measurement data \cite{naderi2020approaching}. And it validates such attacks will not only mislead the optimized operation strategy but also reduce the reliability of the whole energy system \cite{naderi2021hardware}.
In this case, another direction is how to design coordinated remedial action schemes in energy markets
e.g., Ref.~\cite{naderi2021experimental} to further mitigate impacts of this risk and reduce economic losses.

\end{document}